\documentstyle[aps,twocolumn,prl,epsfig]{revtex}

\begin{document}

\title{         
Analysis of the 
Knight shift data on Li and Zn substituted 
YBa$_2$Cu$_3$O$_{6+x}$
}

\author{ N.\ Bulut }

\address{
Department of Physics, Ko\c{c} University, Sariyer, 80910 Istanbul, Turkey \\
nbulut@ku.edu.tr} 

\draft

\twocolumn[\hsize\textwidth\columnwidth\hsize\csname@twocolumnfalse\endcsname
\maketitle 

\begin{abstract}
The Knight shift data on Li and Zn substituted 
YBa$_2$Cu$_3$O$_{6+x}$
are analysed using an itinerant model with short-range 
antiferromagnetic correlations.
The model parameters,  
which are determined by fitting the experimental data
on the transverse nuclear relaxation rate $T_2^{-1}$ of pure
YBa$_2$Cu$_3$O$_{6+x}$,
are used to calculate the Knight shifts for various nuclei 
around a nonmagnetic impurity located in the CuO$_2$ planes.
The calculations are carried out for Li and Zn impurities substituted 
into optimally doped and underdoped 
YBa$_2$Cu$_3$O$_{6+x}$.
The results are compared with the $^7$Li and $^{89}$Y
Knight shift measurements on these materials.
 
\end{abstract}


]


\section{Introduction}

The substitution of nonmagnetic impurities 
into cuprates serves as a probe 
of the electronic correlations in these materials.
In particular, it has been found that 
the nonmagnetic impurities 
significantly influence the magnetic correlations in the 
normal state
\cite{Mahajan94,Bobroff,Mendels,Mahajan99,Alloul99,MacFarlane,Julien,Ishida93,Zheng,Ishida96}.
The $^{89}$Y NMR experiments \cite{Mahajan94,Mahajan99}
find that when Zn is substituted into 
YBa$_2$Cu$_3$O$_{6+x}$, 
the Knight shifts 
of the Y sites near the impurity
are strongly enhanced as the temperature is lowered. 
It has been also found that the Li impurities 
induce changes in the local magnetic environment 
similar to that of Zn \cite{Bobroff}.
A Curie-like 
$T$ dependence is found in the SQUID 
measurements of the uniform susceptibility for 
Zn substituted 
YBa$_2$Cu$_3$O$_{6+x}$ \cite{Mendels}
as well.
It has been also shown that the Zn impurities 
induce a Curie-like $T$ dependence in the broadening
of the $^{63}$Cu(2) spectral line in 
YBa$_2$Cu$_3$O$_{6.7}$ \cite{Julien}.
The effects of nonmagnetic impurities 
in YBa$_2$Cu$_3$O$_{6+x}$
have been also 
probed by the measurements of 
the $^{89}$Y and $^7$Li NMR relaxation rates
$T_1^{-1}$ \cite{Mahajan94,Mahajan99,MacFarlane}
and by the inelastic neutron scattering experiments
\cite{Kakurai,Sidis,Fong,Sidis2000}.
The $T_1^{-1}$ measurements have been also carried out 
for Cu(2) in Zn substituted 
YBa$_2$Cu$_3$O$_{7}$ 
\cite{Ishida93}
and YBa$_2$Cu$_4$O$_{8}$ \cite{Zheng} systems.
The effects of the nonmagnetic impurities on the magnetic 
correlations were also studied in 
La$_{1.85}$Sr$_{0.15}$CuO$_4$ with Al impurities 
\cite{Ishida96}.

Numerous theoretical studies 
\cite{Heisenberg,Sandvik,Poilblanc,Gabay,Li,Bulut2000,neutron}
have been carried out for exploring 
the effects of the nonmagnetic impurities on the magnetic 
properties of the cuprates.
The effects of a nonmagnetic impurity embedded into 
the 2D Heisenberg lattice \cite{Heisenberg,Sandvik}
as well as in gapped Heisenberg antiferromagnets \cite{Sandvik}
have been studied.
The exact diagonalization calculations have been used to 
study the effects of a nonmagnetic impurity in the 
$t$-$J$ model \cite{Poilblanc}.
Theoretical studies have been also carried out for the
underdoped 
YBa$_2$Cu$_3$O$_{6+x}$,
where there is a spin pseudogap 
\cite{Gabay}.
Calculations of the neutron scattering intensity for Zn substituted 
YBa$_2$Cu$_3$O$_{6+x}$ 
have been also carried out \cite{Li,neutron}.

In this paper, a simple framework \cite{Bulut2000}
of one nonmagnetic impurity 
embedded into the lattice of the two-dimensional (2D) Hubbard 
model will be used to make comparisons with the $^7$Li
and $^{89}$Y Knight shift measurements on the 
Li and Zn substituted 
YBa$_2$Cu$_3$O$_{6+x}$.
The purpose of this paper 
is to study the effects of the nonmagnetic impurities 
on the magnetic correlations in the normal state of the cuprates. 
In the following, the 2D Hubbard model and the random-phase
approximation (RPA) will be used for modelling 
the antiferromagnetic correlations in pure
YBa$_2$Cu$_3$O$_{6+x}$.
The 2D Hubbard model is defined by 
\begin{eqnarray}
\label{Hubbard}
H=-t\sum_{\langle i,j\rangle ,\sigma} 
(c^{\dagger}_{i\sigma}c_{j\sigma}
+c^{\dagger}_{j\sigma}c_{i\sigma})
+ U \sum_i 
c^{\dagger}_{i\uparrow} c_{i\uparrow} 
c^{\dagger}_{i\downarrow} c_{i\downarrow} \nonumber \\
-\mu \sum_{i,\sigma}
c^{\dagger}_{i\sigma} c_{i\sigma}.
\end{eqnarray}
Here $c_{i\sigma}$ ($c^{\dagger}_{i\sigma}$)
annihilates (creates) an electron with spin $\sigma$
at site $i$,
$t$ is the near-neighbor hopping matrix element,
$U$ is the onsite Coulomb repulsion, and $\mu$ is the 
chemical potential.
The model parameters for the pure system will be 
determined by fitting the experimental data 
\cite{Pennington,Itoh,Imai,Takigawa94}
on the transverse nuclear relaxation rate
$T_2^{-1}$ of pure
YBa$_2$Cu$_3$O$_{6+x}$.

Next, a static extended impurity potential will be used to incorporate 
the effects of one nonmagnetic impurity embedded into the 2D
Hubbard model.
There is considerable support for using an extended
impurity potential for modelling the effective impurity potential 
\cite{Xiang,Ziegler}.
Here, it will be assumed that the effective interaction between an impurity
located at site ${\bf r}_0$ and a quasiparticle at ${\bf r}$ 
can be approximated by the static potential
\begin{equation}
\label{Veff}
V_{\rm eff}({\bf r}_0,{\bf r}) = 
V_0 \delta({\bf r}_0,{\bf r}) + 
V_1 \sum_{\alpha=1}^4 \delta({\bf r},{\bf r}_0 + \rho_{\alpha}),
\end{equation}
which has a range of one lattice spacing.
In Eq.~(\ref{Veff}), 
$\alpha$ sums over the four near-neighbor sites of the 
impurity as illustrated in Fig.~1.
Using this simple form of $V_{\rm eff}$, 
the Knight shift of various nuclei around the impurity will be 
calculated, and the results will be compared with the experimental 
data.

In the following, 
since the Zn and Li impurities are considered to have closed outer 
electronic shells with $S=0$, 
the onsite component $V_0$ will be set to 
a large negative value.
On the other hand, the near-neighbor component $V_1$ will 
be used as a free model parameter. 
Clearly, this form of $V_{\rm eff}$, Eq.~(\ref{Veff}), is a simple 
approximation. 
For instance, the range of $V_{\rm eff}$ 
could be more than one lattice spacing.
Furthermore, $V_{\rm eff}$ could have scattering in the magnetic channel
\cite{Hirschfeld}.
In fact, any magnetic scattering could 
drastically change the results presented here.
Nevertheless, it is interesting to explore the consequences of this form 
of $V_{\rm eff}$ on the Knight shifts within this simple framework.
Within this model at the level of RPA, 
the fitting of the $^7$Li and 
$^{89}$Y Knight shift data will require that $V_1$
is weakly attractive.

The comparisons will be first carried out
with the $^7$Li Knight shift, $^7K$, measurements in Li
substituted
YBa$_2$Cu$_3$O$_{6.97}$
by Bobroff {\it et al.} \cite{Bobroff}.
Here, $^7K$ will be calculated 
for various values of $V_1$.
It will be seen that the temperature dependence of $^7K$ 
can be fitted by using a $V_1\approx -0.15t$. 
The effects of the nonmagnetic impurity on the 
$^{89}$Y and the $^{63}$Cu(2) Knight shifts for the sites near the 
impurity will also be given for 
this compound.

According to this simple model \cite{Bulut2000}, 
the anomalous $T$ dependence 
observed in the Knight shift data is due 
to the enhancement of the antiferromagnetic correlations 
in the local environment of the impurity. 
In this model, 
the scattering of the antiferromagnetic spin fluctuations
by the impurity with large momentum transfers near $2{\bf k}_F$, 
where ${\bf k}_F$ is the fermi wave vector, 
plays a particularly important role
\cite{Bulut2000,neutron}.
These scatterings locally enhance the antiferromagnetic 
correlations, and, in addition, allow for the Knight shift 
of the nuclear sites near the impurity to be coupled to the 
locally 
enhanced antiferromagnetic correlations.

Because of the presence of the spin pseudogap in 
underdoped
YBa$_2$Cu$_3$O$_{6+x}$,
it is indeed difficult 
to extend this analysis to this compound.
However, the Knight shift experiments on underdoped 
YBa$_2$Cu$_3$O$_{6+x}$ 
yielded interesting results
on the real-space structure of the induced magnetic 
correlations around the impurity.
These measurements have been carried out for 
$^7$Li in Li substituted 
YBa$_2$Cu$_3$O$_{6.6}$ \cite{Bobroff} and for 
$^{89}$Y in Zn or Li substituted 
YBa$_2$Cu$_3$O$_{6.64}$ \cite{Mahajan94,Mahajan99}.
For $^{89}$Y, two magnetic resonance satellites 
in addition to the main line are found in the 
presence of the Zn or Li impurities in underdoped 
YBa$_2$Cu$_3$O$_{6+x}$.
These satellites have been identified as belonging to the 
first and the second neighbour Y sites of the impurity. 
The availability of the Knight shift data on $^7$Li and 
the two $^{89}$Y satellites gives valuable information on
the real-space structure of the magnetic correlations 
in the local environment 
of the impurity \cite{Mahajan99}. 
The $^7$Li and $^{89}$Y Knight shift 
measurements by Mahajan {\it et al.} 
\cite{Mahajan94,Mahajan99} 
and the data on the Knight shift of $^{63}$Cu(2), 
$^{63}K_c$, by Julien {\it et al.} 
\cite{Julien}
point out that a staggered 
polarisation cloud forms around the impurity
when a uniform magnetic field is applied. 
Hence, here, using simple assumptions, the 
results of the calculations will be compared with the 
$^7$Li and $^{89}$Y Knight shifts in order to 
see whether this simple model has some 
of the features observed experimentally.
Results on $^{63}K_c$ will also be presented. 
It is important to keep in mind that the purpose 
in this paper will
not be to present a theory of the magnetic correlations 
in the pseudogap state within the presence of nonmagnetic 
impurities.
This is clearly beyond the simple model used here. 
Rather, the purpose is to study the real-space structure 
of the deformations induced by the impurity
by making simple assumptions. 
It would have been more desirable to carry out such a 
study for optimally doped 
YBa$_2$Cu$_3$O$_{7}$,
where this model is more applicable,
but in that case the $^{89}$Y lines are not resolved. 
Here, the model parameters determined by fitting the $T_2^{-1}$ 
data on pure 
YBa$_2$Cu$_3$O$_{6.63}$
will be used to calculate the $^7$Li and $^{89}$Y
Knight shifts. 
Initially, the Knight shifts will be calculated 
by neglecting the opening of the pseudogap in the 
${\bf q}\rightarrow 0$ component of the 
magnetic susceptibility of the pure system. 
Even in this case, it will be found that for 
$V_1$ between $-0.125t$ and 
$-0.15t$, the real-space pattern of the 
magnetic correlations around the impurity 
is similar to what is seen experimentally.
Later on, a pseudo gap in the 
${\bf q}\rightarrow 0$ component of the 
magnetic susceptibility will be introduced artificially. 
In this case, comparisons with the data require
$V_1\approx -0.15t$.
It has to be re-emphasized that the results on the 
underdoped 
YBa$_2$Cu$_3$O$_{6+x}$
needs to be interpreted cautiously.
The treatment of the pseudogap is clearly not rigorous.
An especially important point is that the presence of the 
pseudogap can introduce new physics which is quite different 
than the simple scenario discussed here using 
a weak-coupling model.

In Section II below, the model will be introduced.
After comparisons with the experimental data 
in Sections III.A and III.B, 
the pattern of the magnetic correlations around the impurity 
will be shown in Section III.C.
The implications of these calculations and the role of the 
antiferromagnetic correlations will be discussed 
in Section IV. 
In addition, here, 
the dependence of the results on the 
effective bandwidth will be studied, 
and the hyperfine couplings used in the calculations will be 
compared with the experimental estimates.
In Section V, the summary and the conclusions will be given.

\section{Model}

Here, the framework for calculating the 
Knight shifts will be introduced.
In Section II.A, the modelling of the magnetic susceptibility 
for pure 
YBa$_2$Cu$_3$O$_{6+x}$
will be presented.
The model parameters, which will be used to parameterise
the antiferromagnetic correlations of pure
YBa$_2$Cu$_3$O$_{6+x}$,
will be determined by fitting the experimental 
data on the transverse nuclear relaxation rate 
$T_2^{-1}$ of 
YBa$_2$Cu$_3$O$_{6+x}$ \cite{Pennington,Itoh,Imai,Takigawa94}.
In Section II.B, the static extended impurity potential 
used for approximating the interaction between the 
impurity and the electrons will be discussed.
In the following section, the calculation of the magnetic 
susceptibility for the case of one impurity will be presented. 
In Section II.D, the hyperfine interactions for the $^7$Li and $^{89}$Y 
nuclear spins will be introduced.

\subsection{Parameterisation of $\chi({\bf q},\omega)$ 
for pure YBa$_2$Cu$_3$O$_{6+x}$}

In this paper, it will be assumed that the magnetic correlations of pure 
YBa$_2$Cu$_3$O$_{6+x}$
can be approximated by the 
two-dimensional Hubbard model.
The magnetic susceptibility of the pure system is defined by 
\begin{equation}
\chi_{pure}({\bf q})= \int_0^{\beta} d\tau \,
\langle m^-({\bf q},\tau) m^+({\bf q},0) \rangle.
\end{equation}
Here, $\tau$ is the Matsubara time, 
$m^+({\bf q},0)=N^{-1/2} \sum_{\bf p}
c_{{\bf p}+{\bf q}\uparrow} c_{{\bf p}\downarrow}$,
$m^-({\bf q},\tau)=
e^{H\tau} m^-({\bf q},0) e^{-H\tau}$,
and 
$m^-({\bf q},0)=N^{-1/2} \sum_{\bf p}
c_{{\bf p}+{\bf q}\downarrow} c_{{\bf p}\uparrow}$.
The RPA form for $\chi_{pure}$ is 
\begin{equation}
\label{chirpa}
\chi_{pure}({\bf q}) = { \chi_0^L({\bf q}) \over 1-U \chi_0^L({\bf q})}.
\end{equation}
Here,
$U$ is the renormalized Coulomb repulsion 
and the static Lindhard susceptibility 
$\chi_0^L({\bf q})$ is given by
\begin{equation}
\label{Lindhard}
\chi_0^L({\bf q}) = {1\over N}
\sum_{\bf p} 
{ f(\varepsilon_{{\bf p}+{\bf q}}) - f(\varepsilon_{\bf p}) 
\over 
\varepsilon_{\bf p}
- \varepsilon_{{\bf p}+{\bf q}} },
\end{equation} 
where $f(\varepsilon_{\bf p})$ is the fermi factor 
and 
$\varepsilon_{\bf p}=-2t(\cos{p_x} + \cos{p_y})-\mu$.
Using this form for $\chi_{pure}({\bf q})$ with the renormalised 
values of $U$ and the hopping matrix element $t$,
the longitudinal and the transverse relaxation rates 
$T_1^{-1}$ and $T_2^{-1}$ for 
YBa$_2$Cu$_3$O$_{7}$ have been calculated \cite{T1,T2}.
In this model, the renormalization of $U$ 
is due to the particle-particle correlations and the single-particle 
self-energy corrections, and the renormalization of the 
bandwidth is due to the Coulomb correlations \cite{PhysicaC}.
The value of $U$ will be determined by fitting the experimental 
data on the $^{63}$Cu(2) transverse relaxation rate $T_2^{-1}$ 
in pure 
YBa$_2$Cu$_3$O$_{6+x}$.
In the following, the effective bandwidth $W$
will be taken to be $1$ eV
as in Refs.~\cite{T1,T2}.
Later, in Section IV.C the dependence of the results 
on $W$ will be studied.

The transverse nuclear relation rate $T_2^{-1}$ of $^{63}$Cu(2) 
for an orienting field along the $c$-axis is given by 
\cite{Pennington}
\begin{eqnarray}
\label{T2}
\bigg({1 \over T_2}\bigg)^2 = 
{0.69 \over 32\hbar^2 }
\bigg\{
{1\over N}\sum_{{\bf q}} |A({\bf q})|^4\chi_{pure}^2({\bf q}) \nonumber \\
- 
\bigg(
{1\over N}\sum_{{\bf q}} |A({\bf q})|^2 \chi_{pure}({\bf q})
\bigg)^2 
\bigg\}
\end{eqnarray}
where 
\begin{equation}
A({\bf q}) =A_c + 2B(\cos{q_x}+\cos{q_y}).
\end{equation}
Here, the following hyperfine interaction \cite{MR}
has been assumed for 
the $^{63}$Cu nuclear spin at site $i$,
\begin{equation}
\label{MR}
\sum_{\alpha=x,y,z} A_{\alpha \alpha} 
I_i^{\alpha} S_i^{\alpha}  + 
B \sum_{\delta=1}^4 
{\bf I}_i \cdot {\bf S}_{i+\delta},
\end{equation}
where $\delta$ sums over the four nearest neighbours of site $i$.
For the hyperfine couplings, 
$A_c=-4B=-2.45\times 10^{-18}$ erg,
corresponding to 164 kOe/$\mu_B$,
will be used as in Ref. \cite{T2}.
In fitting the $T$ dependence of $T_2^{-1}$ with the RPA
form of Eq.~(\ref{chirpa}), $U$ will be adjusted
as $T$ is varied.
Clearly, this is a simple, approximate procedure for
determining the strength of the antiferromagnetic 
correlations in the pure system.
In fact, the model is so simple that electron filling 
$\langle n\rangle$ will be taken to be 0.86 for both 
the underdoped and optimally doped systems. 
Hence, the only feature which differentiates between the 
underdoped and the optimally doped systems in this model is 
the strength of the antiferromagnetic correlations 
of the pure system which
is set by adjusting $U$
to fit the $T_2^{-1}$ data.
As discussed in Ref.\cite{Bulut2000}, the antiferromagnetic 
correlations of the impure system can get significantly 
enhanced by the impurity scattering as compared 
to those in the pure system.

The fitting of $T_2^{-1}$ will be first carried out for 
optimally doped
YBa$_2$Cu$_3$O$_{6+x}$.
The transverse relaxation rate $T_2^{-1}$ of 
$^{63}$Cu(2) 
has been measured by Pennington and Slichter \cite{Pennington}
in YBa$_2$Cu$_3$O$_{7}$, 
and by Imai {\it et al.} \cite{Imai}
in YBa$_2$Cu$_3$O$_{6.9}$.  
The data by Imai {\it et al.} \cite{Imai}
are represented by the open circles 
in Fig.~2(a).
Here, it is seen that $T_2^{-1}$ increases linearly as $T$ is
lowered from 300 K to 100 K.
The dashed line in this figure has been obtained by using the 
RPA form of Eq.~(\ref{chirpa}) in Eq.~(\ref{T2}) and by adjusting 
$U$ to fit the experimental data.
The resultant values of $U(T)$ are given by the dashed line 
in Fig.~2(b).
Here one observes that as $T$ is lowered from 300 K to 100 K, 
$U(T)$ decreases by 5\%.
Note also that the dashed curve in Fig.~2(a) has been 
extrapolated up to 400 K, since in Section III the Knight shifts will
be calculated up to this temperature.

The filled circles in Fig. 2(b) represent the 
$T_2^{-1}$ data for $^{63}$Cu(2) in 
YBa$_2$Cu$_3$O$_{6.63}$ 
measured by Takigawa \cite{Takigawa94}.
Here it is seen that $T_2^{-1}$ varies almost linearly 
for $T$ between 160 K and 300 K.
Below 160 K, $T_2^{-1}$ saturates.
Because of this, the fitting of 
$T_2^{-1}$ for $T< 160$ K will require more 
attention.
This saturation might be due to two reasons:
(1) the suppression of $\chi_{pure}({\bf q}\sim 0)$ because of the 
opening of the magnetic pseudogap,
or
(2) the saturation of $\chi_{pure}({\bf q})$ for 
${\bf q}\sim (\pi,\pi)$.
In the first case, 
$\chi_{pure}({\bf q}\sim (\pi,\pi))$ would continue
to increase as $T$ is lowered and the saturation 
of $T_2^{-1}$ would be due directly to the 
suppression of $\chi_{pure}({\bf q}\sim 0)$.
In the other case, 
the saturation of $T_2^{-1}$ would be due
to the saturation of the antiferromagnetic correlations.
In order to take into account these two possibilities, the fitting 
of $T_2^{-1}$ below 160 K will be done in two ways.
In the first case, 
the saturation of $T_2^{-1}$ will be attributed to the 
suppression of $\chi_{pure}({\bf q}\sim 0)$ and it will be 
assumed that without this effect $T_2^{-1}$ would have 
continued to increase linearly as $T$ is lowered below 160 K.
Hence, in this case, $T_2^{-1}$ will be extrapolated for 
$T<160$ K as shown by the solid line in Fig.~2(a).
The resultant values of $U(T)$ are shown by the 
solid curve in Fig.~2(b).
In the second case, the $T_2^{-1}$ data below 
160 K will be fitted directly as shown 
by the dotted curve in Fig.~2(a).
The resultant $U(T)$ is shown by the dotted curve
in Fig.~2(b).
By comparing the solid and the dotted curves for $T<160$ K in 
Figs.~2(a) and (b), it is seen that small changes 
in $U$ produce significant
changes in $T_2^{-1}$.
This is because the system is close to a magnetic instability with
a large Stoner enhancement for 
${\bf q}\sim (\pi,\pi)$.

At this point, it is useful to see the strength of the 
antiferromagnetic correlations required for fitting 
the $T_2^{-1}$ data.
The dashed curve in Fig.~3(a) shows $\chi_{pure}({\bf q})$ 
versus ${\bf q}$ obtained 
by fitting the $T_2^{-1}$ data on pure
YBa$_2$Cu$_3$O$_{6.9}$
at 100 K.
The solid and the dotted curves represent 
$\chi_{pure}({\bf q})$ obtained 
with the first and the second scenarios, respectively, 
for fitting $T_2^{-1}$ of 
YBa$_2$Cu$_3$O$_{6.63}$ 
at 100 K \cite{Takigawa94}.
For the form of $\chi_{pure}({\bf q})$ used here,
Eqs.~(\ref{chirpa}) and (\ref{Lindhard}), the peak in 
$\chi_{pure}({\bf q})$ occurs at an incommensurate wave vector
${\bf Q}^*$ away from ${\bf Q}=(\pi,\pi)$.
The neutron scattering experiments \cite{Arai,Bourges} on
YBa$_2$Cu$_3$O$_{6+x}$
find that in the superconducting state, 
the spin-fluctuation spectral weight 
${\rm Im}\,\chi_{pure}({\bf q},\omega)$ 
peaks away from $(\pi,\pi)$, but with 
an incommensuration which is more than what is 
observed in Fig.~3(a).
The form of $\chi_{pure}({\bf q})$ used here is clearly a crude
approximation.
The results shown below will not depend sensitively on the amount of 
the incommensuration, rather they will depend on the 
total integrated weight 
in the ${\bf q}\sim (\pi,\pi)$ region 
of the ${\bf q}$-space.
Further results on the nature of the antiferromagnetic 
correlations are shown in Fig.~3(b).
The dashed line in Fig.~3(b) shows $\chi_{pure}({\bf Q}^*)$
versus $T$ for 
optimally doped
YBa$_2$Cu$_3$O$_{6+x}$.
Also shown in this figure by the solid  
line are the results for $\chi({\bf Q}^*)$
of the impure system
with 0.5\% randomly distributed impurities
which will be discussed below in Section IV.A.
Fig.~3(c) shows 
similar results for 
$\chi({\bf Q}^*)$ of underdoped 
YBa$_2$Cu$_3$O$_{6+x}$.
Here, it is seen that $\chi({\bf Q}^*)$ gets significantly 
enhanced by the substitution of the nonmagnetic impurities, 
especially for underdoped 
YBa$_2$Cu$_3$O$_{6+x}$.
By comparing Fig.~2(a) with Figs.~3(b) and (c), 
it is also seen that
the temperature dependence
of $T_2^{-1}$ and $\chi_{pure}({\bf Q}^*)$ are 
closely related, as expected.

\subsection{Effective Impurity Potential}

The effects of one nonmagnetic impurity located at site ${\bf r}_0$
will be taken into account by adding to Eq.~(\ref{Hubbard}) 
the term 
\begin{equation}
\label{Vimp}
\sum_{i,\sigma} V_{\rm eff}({\bf r}_0,{\bf r}_i) 
c_{i\sigma}^{\dagger} c_{i\sigma}
\end{equation}
where $V_{\rm eff}({\bf r}_0,{\bf r})$
is given by Eq.~(\ref{Veff}).
The importance of using an extended impurity potential has been
previously noted {\cite{Xiang,Ziegler}.
In Ref.~\cite{Ziegler}, it has been pointed out that 
the Coulomb correlations of the host could cause the extended 
nature of the effective interaction. 
The fact that the defects generated by the electron irradiation 
of the samples have similar effects on the $T_c$
suppression \cite{Tc} and on the magnetic 
properties \cite{irradiation}
also supports using an impurity potential 
with only a potential scattering term.

In the following, the onsite component of the impurity
potential, $V_0$, will be set to a large negative value, $-100t$,
in order to model the closed electronic 
shell of the nonmagnetic impurity.
As long as $|V_0|$ has a large value, 
its exact magnitude or its sign does not play an important role.
For instance, using $V_0=-200t$ instead of $-100t$ 
does not change the results.
The near-neighbour component $V_1$, on the other hand, 
will be used as a free parameter in fitting the Knight shift 
measurements on Li and Zn substituted 
YBa$_2$Cu$_3$O$_{6+x}$.
In the next section, 
the effects of this static extended impurity potential 
on the magnetic susceptibility 
will be calculated.

\subsection{Magnetic susceptibility within the presence 
of one nonmagnetic impurity}

The calculation of $\chi$ within the presence of a non-magnetic 
impurity used here follows that given in Ref.~\cite{Bulut2000}.
First, the effects of the impurity on the single-particle 
Green's function will be calculated.
The single-particle Green's function in Matsubara frequency 
space is defined by 
\begin{equation}
G({\bf r}_i,{\bf r}_j,i\omega_n) = 
- \int_0^{\beta} d\tau \, 
e^{i\omega_n \tau}
\langle c_{i\sigma}(\tau) 
c^{\dagger}_{j\sigma}(0) \rangle,
\label{G}
\end{equation}
where $\omega_n=(2n+1)\pi T$.
For the pure system with $U=0$, 
one has 
\begin{equation}
G_0({\bf r}_i,{\bf r}_j,i\omega_n) = 
{1\over N}
\sum_{\bf p} 
e^{ i {\bf p} \cdot ({\bf r}_i-{\bf r}_j) }
G_0({\bf p},i\omega_n)
\end{equation}
where
\begin{equation}
G_0({\bf p},i\omega_n)
={1 \over i\omega_n-\varepsilon_{\bf p} }.
\end{equation}
If an impurity is introduced at site ${\bf r}_0$,
then one gets 
\begin{eqnarray}
\label{GT}
&&G({\bf r},{\bf r'},i\omega_n) = 
G_0({\bf r},{\bf r'},i\omega_n)  \nonumber \\
&& + 
\sum_{{\bf r}'',{\bf r}'''} 
G_0({\bf r},{\bf r}'',i\omega_n)  
T({\bf r}'',{\bf r}''',i\omega_n)
G({\bf r}''',{\bf r}',i\omega_n), 
\end{eqnarray}
where the $T$-matrix for the impurity scattering is given by 
\begin{eqnarray}
\label{Tmatrix}
T({\bf r},&&{\bf r'},i\omega_n)=
\delta({\bf r},{\bf r'}) 
V_{\rm eff}({\bf r}_0,{\bf r}) \nonumber \\
&&+ 
\sum_{{\bf r''}} V_{\rm eff}({\bf r}_0,{\bf r})
G_0({\bf r},{\bf r''},i\omega_n) 
T({\bf r''},{\bf r'},i\omega_n). 
\end{eqnarray} 
The calculation of $G({\bf r},{\bf r'},i\omega_n)$ 
is illustrated diagrammatically in Fig.~4(a).

When the translational invariance is broken,
the magnetic susceptibility is defined in real space as
\begin{equation}
\chi({\bf r},{\bf r'},i\omega_m) = 
\int_0^{\beta} d\tau \, 
e^{i\omega_m \tau}
\langle 
m^{-}({\bf r},\tau)  m^{+}({\bf r'},0) 
\rangle,
\label{chi}
\end{equation}
where $m^{+}(\bf r)=c^{\dagger}_{{\bf r}\uparrow} 
c_{{\bf r}\downarrow}$, and
$m^{-}({\bf r})=c^{\dagger}_{{\bf r}\downarrow} 
c_{{\bf r}\uparrow}$.
The effects of a single impurity 
will be first calculated for $U=0$,
giving the irreducible susceptibility 
$\chi_0({\bf r},{\bf r'},i\omega_m)$, 
and 
then the effects of the Coulomb correlations will 
be included.
The diagrams representing the effects of the impurity 
on $\chi_0$
are shown in Figs.~4(b) and (c).
Both the self-energy and the vertex corrections 
need to be included \cite{Langer}, and the 
resulting expression for 
$\chi_0({\bf r},{\bf r}',i\omega_m)$ 
is given by 
\begin{eqnarray}
\label{chi0}
\chi_0({\bf r},{\bf r'},i\omega_m) = - T &&\sum_{i\omega_n}
\bigg[ 
G({\bf r},{\bf r}', i\omega_{n}+i\omega_m) \nonumber \\
\times 
G_0({\bf r}',{\bf r},i\omega_n) 
+
G_0&&({\bf r},{\bf r}',i\omega_{n}+i\omega_m) 
G({\bf r}',{\bf r},i\omega_n) \nonumber \\
-
G_0&&({\bf r},{\bf r}',i\omega_{n}+i\omega_m)
G_0({\bf r}',{\bf r},i\omega_{n}) 
\bigg] \nonumber \\
- T \sum_{i\omega_n} 
\sum_{{\bf r}_1,{\bf r}_2,{\bf r}_3,{\bf r}_4}
G_0({\bf r},&&{\bf r}_1, i\omega_{n}+i\omega_m) \nonumber \\  
\times
G_0({\bf r}_2,{\bf r}',i\omega_{n}+i&&\omega_m) 
T({\bf r}_1,{\bf r}_2,i\omega_{n}+i\omega_m) \nonumber \\
\times 
T({\bf r}_3,{\bf r}_4,i\omega_n) 
G_0&&({\bf r}_3,{\bf r},i\omega_n) 
G_0({\bf r}',{\bf r}_4,i\omega_n). 
\end{eqnarray}
Note that the irreducible impurity-scattering 
vertex has been used 
in calculating the vertex corrections to $\chi_0$
instead of the reducible one.
This is necessary in order to prevent double counting, 
since here the effects of only one impurity 
are calculated.

Next,
the Coulomb correlations are included 
by solving the RPA equation
\begin{eqnarray} 
\chi({\bf r},{\bf r}',i\omega_m) &&= 
\chi_0({\bf r},{\bf r}',i\omega_m) \nonumber \\
&&+ 
U \sum_{{\bf r}''}
\chi_0({\bf r},{\bf r}'',i\omega_m)
\chi({\bf r}'',{\bf r}',i\omega_m)
\label{rpa}
\end{eqnarray}
for $\chi({\bf r},{\bf r}',i\omega_m)$.
Here $\chi$ is calculated for an $N$-site
square lattice with one nonmagnetic impurity located 
at the center using periodic boundary conditions.
The calculations are carried out 
on sufficiently large lattices so that the 
finite size effects are small.
In Ref.~\cite{Bulut2000}, the finite size effects on 
$\chi$ have been studied in detail.

The Knight shift is determined by 
$\chi({\bf r},{\bf r}')=
\chi({\bf r},{\bf r}',i\omega_m=0)$.
In the following section, the hyperfine 
interactions, which relate 
$\chi({\bf r},{\bf r}')$ to the Knight shifts, will be discussed 
for $^7$Li and $^{89}$Y.

\subsection{Hyperfine interactions for 
$^7$Li and $^{89}$Y}

The hyperfine interaction between the $^7$Li nuclear spin
with $^7I=3/2$ substituted into a Cu(2) site
and the electronic spins will be modelled
by the following hyperfine coupling \cite{Bobroff}
\begin{equation}
C \sum^4_{\delta=1} \,
{^7{\bf I}}\cdot {\bf S}_{\delta}.
\label{A7}
\end{equation}
Here $\delta$ sums over the four Cu(2) sites neighbouring the Li impurity.
The nature of this interaction is similar to that of the transferred
hyperfine coupling of the $^{63}$Cu(2) nuclear spins to the 
electronic spins at the neighbouring Cu(2) 
sites given by the second term
in Eq.~(\ref{MR}) \cite{MR}.
In Section III, 
the magnitude of the $^7$Li 
hyperfine coupling $C$ will be taken to be 
$1.8\times 10^{-20}$ erg,
corresponding to 0.85 kOe/$\mu_B$. 
This choice for the value of $C$ will be discussed later in 
Section IV.C.
As a result of this coupling, Eq.~(\ref{A7}), 
the Knight shift of $^7$Li is (see Appendix I)
\begin{equation}
^7K = {1 \over 2} 
\left( { \gamma_e \over ^7\gamma_n } \right)
C \,
4k({\bf r}=(1,0)),
\label{K7}
\end{equation}
where $k({\bf r})$ is defined by 
\begin{equation}
\label{kr}
k({\bf r}) = \sum_{{\bf r}'}
\chi({\bf r},{\bf r}'),
\end{equation}
and $\gamma_e$ and $^7\gamma_n$ are the gyromagnetic 
ratios of the electron and the 
$^7$Li nuclear spin. 
The factor of four in Eq.~(\ref{K7}) is 
because the Li impurity has four Cu(2) neighbours.

The hyperfine interaction for the $^{89}$Y
nuclear spin with $^{89}I=1/2$ is 
\begin{equation}
D\sum_{\alpha=1} ^8 \,
^{89}{\bf I}\cdot {\bf S}_{\alpha}
\end{equation}
where $\alpha$ sums over the eight Cu(2) sites neighbouring the 
$^{89}$Y nuclear spin.
In Section III, the hyperfine interaction 
$D$ will be assumed to be 
$-2.2\times 10^{-20}$ erg,
corresponding to $-1.0$ kOe/$\mu_B$, 
in order to fit the normal state value of $^{89}K$ in pure 
YBa$_2$Cu$_3$O$_{6+x}$ 
using a bandwidth of $W=1$ eV. 
Later in Section IV.C, the dependence of the results on 
$W$ and on the values of the hyperfine couplings $C$ and $D$
will be discussed. 
Here, it will also be assumed that the $^{89}$Y hyperfine couplings 
do not change upon the substitution of the nonmagnetic 
impurity. 
Then the Knight shift
for $^{89}$Y at a particular Y lattice site is given by 
\begin{equation}
^{89}K = {1 \over 2} 
\left( { \gamma_e \over ^{89}\gamma_n } \right)
D 
\sum_{\alpha=1}^8
k({\bf r}_{\alpha}),
\end{equation}
where ${\bf r}_{\alpha}$ are the locations of the
eight nearest-neighbour Cu(2) sites of the 
Y site in the CuO$_2$ bilayer.
For dilute Li or Zn impurities, the $^{89}$Y nuclear spin will be 
affected by an impurity located in one of the layers of the 
CuO$_2$ bilayer in 
YBa$_2$Cu$_3$O$_{6+x}$.
Here, it will be assumed that the 
changes induced in $^{89}K$ are due only to the changes in 
the magnetic correlations in the layer which contains the 
impurity as it was done in Ref.~\cite{Mahajan99}.
Also, 
in the following, the quantity of interest for $^{89}$Y
will be the change in the Knight shift of $^{89}$Y due 
to the presence of the impurity.
Hence, the calculations will be carried out for 
\begin{equation}
\label{dK89}
\delta^{89} K = 
{1 \over 2} 
\left( { \gamma_e \over ^{89}\gamma_n } \right)
D
\bigg[ \sum_{\alpha=1}^4 k({\bf r}_{\alpha}) - 
4\chi_{pure}\bigg],
\end{equation}
where $\alpha$ sums over the four Cu(2) sites neighbouring 
the impurity in the layer which contains the impurity.
In Eq.~(\ref{dK89}),
$\chi_{pure}$ is the uniform static susceptibility
of the pure system.
The value of $^{89}K$ will depend strongly 
on the location of the $^{89}$Y nuclear spin with respect to the 
impurity.
In Fig.~5, a sketch of the various Y sites with respect to the 
impurity site is given.
Note that the measurement of $^{89}K$ at the Y sites
near the impurity, 
along with that of $^7K$, determines the real-space structure
of $k({\bf r})$ in the environment of the 
impurity. 
For instance, 
$^{89}K$ for Y(1) is determined by 
$k({\bf r}=(1,0))$ and 
$k({\bf r}=(1,1))$, while $^7K$ is set by 
$k({\bf r}=(1,0))$.
In Section~III, the results of the calculations on $^7K$ 
and $^{89}K$ for various Y sites will be shown 
and compared with the experimental data.

\section{Results of the Calculations}

The comparison of the results with the experimental data 
will be carried out first for the optimally doped 
and then for the underdoped
YBa$_2$Cu$_3$O$_{6+x}$.

\subsection{Results on optimally doped 
YBa$_2$Cu$_3$O$_{6+x}$}

The filled circles in Fig.~6(a) denote the experimental 
data by Bobroff {\it et al.} \cite{Bobroff}
on the $^7$Li Knight shift $^7K$ for 
YBa$_2$Cu$_3$O$_{6.97}$ 
with dilute Li impurities.
The curves in this figure represent the results of the calculations
of $^7K$ for various values of $V_1$ as indicated
next to the curves.
The model parameters other than $V_1$ were already set 
by fitting the $T_2^{-1}$ data on pure 
YBa$_2$Cu$_3$O$_{6.9}$ 
by Imai {\it et al.} \cite{Imai} 
as discussed in Section II.A.
Here, 
it is seen that for $V_1\approx -0.15t$ the calculated 
values of $^7K$ are in agreement with the 
experimental data. 
It is useful to compare these results with 
what is expected for the $^7$Li
Knight shift if the magnetic correlations had not 
changed upon the substitution of the impurity 
and had remained the same
as in the pure material.
In Fig.~6(b), $^7K_0$ defined by 
\begin{equation}
^7K_0= {1\over 2}  
\bigg( { \gamma_e \over ^7\gamma_n } \bigg)
C 4 \chi_{pure},
\end{equation}
where $\chi_{pure}$ is the uniform susceptibility 
of the pure system,
is plotted as a function of the temperature. 
The solid line has been obtained from
$\chi_{pure}({\bf q}\rightarrow 0)$ 
given by Eq.~(\ref{chirpa})
for optimally doped 
YBa$_2$Cu$_3$O$_{6+x}$.
The dashed line in Fig.~6(b) has been obtained 
by rescaling the $^{89}$Y Knight shift $^{89}K$ of pure 
YBa$_2$Cu$_3$O$_{7}$ \cite{Alloul89} 
with
\begin{equation}
^7K_0 = {1\over 2} 
\bigg( {^{89}\gamma_n C \over ^7\gamma_n D} \bigg)\,
^{89}K.
\end{equation}
The factor of $1/2$ in this expression is because 
Y has eight Cu(2) near neighbours while Li has four.
In these figures, 
it is also seen that for $V_1=0$, 
corresponding to an onsite impurity potential 
in the unitary limit, 
$^7K$ is enhanced with respect to $^7K_0$ 
but this enhancement is not sufficient to explain the 
experimental data. 
If positive values of $V_1$ are used, then 
$^7K$ gets suppressed with respect to the $V_1=0$ case.

The stoichiometry of the samples is a factor 
which could affect the comparisons with the data.
Note that $U(T)$ used in calculating $^7K$ for 
YBa$_2$Cu$_3$O$_{6.97}$
was determined by fitting the $T_2^{-1}$ data on 
YBa$_2$Cu$_3$O$_{6.9}$.
For 
YBa$_2$Cu$_3$O$_{7}$, 
$T_2^{-1}$ was measured at 100 K, and its value is 
$7.7 \pm {0.6}$ msec$^{-1}$ compared to 
$\sim 9.5$ msec$^{-1}$ for 
YBa$_2$Cu$_3$O$_{6.9}$.
The value of $7.7$ msec$^{-1}$ for 
$T_2^{-1}$ requires a slightly smaller value for $U$ and, 
consequently, 
the fitting of $^7K$ is done using larger values of $V_1$.
If one assumes that the $T$ dependence of $T_2^{-1}$ 
is given by a line which passes through $7.7$ msec$^{-1}$ 
at 100 K and which is parallel to the dashed line in Fig.~2(a), 
then the $^7K$ data can be fitted by using $V_1\approx -0.275t$.

Based on these results and considering the simplicity 
of the model, the conclusions which can be reached 
are limited. 
Probably, the best thing to say is that the analysis of the 
$^7K$ data on optimally doped
YBa$_2$Cu$_3$O$_{6+x}$
within this model at the level of RPA 
requires an impurity potential which is 
weakly attractive at the nearest-neighbor Cu(2) sites.

The results of the calculations for $^{89}K$ are shown
in Fig.~7(a). 
Here,
the temperature dependence 
of $\delta^{89}K$ is shown 
for the first four nearest-neighbor Y sites which are indicated
in Fig.~5.
In obtaining these results, 
$V_1=-0.15t$ was used.
Note that in Fig.~7(a) the expected values of the Knight 
shifts are close to each other. 
Experimentally, for 
optimally doped 
YBa$_2$Cu$_3$O$_{6+x}$, 
the expected resonance lines for the $^{89}$Y nuclei near 
the Zn impurity are not resolved.

Next, results on how the nonmagnetic impurity affects the 
$^{63}$Cu(2) Knight shifts in this model are presented. 
The $^{63}$Cu(2) Knight shift at site ${\bf r}_i$ 
with the orienting magnetic field ${\bf H}||{\bf c}$ 
is given by 
\begin{equation}
\label{Kc}
^{63}K_c({\bf r}_i) = 
{1 \over 2} 
\bigg( {\gamma_e \over ^{63}\gamma_n } \bigg)
\bigg[
A_c k({\bf r}_i) + 
B\sum_{\alpha=1}^{4} k({\bf r}_{\alpha})
\bigg],
\end{equation}
where $\alpha$ sums over the four Cu(2) neighbors of 
${\bf r}_i$.
Figure 7(b) shows the $T$ dependence of 
$^{63}K_c({\bf r}_i)$ for $V_1=-0.15t$
for the first four neighboring Cu(2) sites 
of the impurity, which correspond to 
${\bf r}_i=(1,0)$, (1,1), (2,0), and (2,1).
For the pure system, 
$^{63}K_c$ vanishes since 
$A_c+4B=0$.
Here, it is seen that $^{63}K_c$ 
has a staggered pattern;
it is negative 
at the sublattice of the (1,0) site 
and it is positive at the sublattice of the impurity.
This spatial pattern of 
$^{63}K_c({\bf r}_i)$
is due to the staggered pattern of 
$k({\bf r})$, which will be discussed in Sections III.C
and IV.A.
The measurements of $^{63}K_c$ have been 
carried out in 
YBa$_2$Cu$_3$O$_{6.7}$
by Julien {\it et al.} \cite{Julien}.
While the individual $^{63}$Cu(2) lines
are not resolved, it has been observed that the broadening of the 
linewidth has a Curie-like $T$ dependence. 
This has been attributed 
to the development of a straggered polarization cloud 
around the impurity \cite{Julien}. 

\subsection{Results on underdoped YBa$_2$Cu$_3$O$_{6+x}$ }

The analysis of the Knight shift data on the underdoped 
YBa$_2$Cu$_3$O$_{6+x}$ is considerably more complicated 
because of the magnetic pseudogap.
This is so because the origin of the magnetic pseudogap 
in this material 
is currently an unresolved issue.
Hence, it is necessary to note that the purpose here is not 
to develop a theory for the magnetic susceptibility 
of the underdoped 
YBa$_2$Cu$_3$O$_{6+x}$
with impurities.
Instead, the purpose is to explore any possible role of the
antiferromagnetic correlations in producing the 
anomalous $T$ dependence of the Knight shift data in 
underdoped 
YBa$_2$Cu$_3$O$_{6+x}$
within the presence of nonmagnetic impurities.
The emphasis here will be on the real-space structure 
of the magnetic correlations around the impurity.

The $^7$Li Knight shift measurements by Bobroff {\it et al.}
\cite{Bobroff} on underdoped
YBa$_2$Cu$_3$O$_{6+x}$ 
with Li impurities found that $^7K$ increases rapidly 
as $T$ is lowered.
In Fig.~8(a), the experimental data on
$\delta^7K$ defined by 
\begin{equation} 
\delta^7K = ^7K - ^7K_0
\end{equation}
are compared with the results of the calculations. 
The comparisons are carried out for 
$\delta ^7K$ rather than for $^7K$ in order to compensate 
for the effect of the magnetic pseudogap 
which is not taken into account here.  
The data points shown by the filled circles 
in Fig.~8(a) were obtained 
by subtracting from the $^7K$ data of Bobroff {\it et al.}
\cite{Bobroff} 
the quantity
\begin{equation}
^7K_0 = 
{1 \over 2}
\bigg( {^{89}\gamma_n C \over ^7\gamma_n D} \bigg) \,
^{89}K({\rm main}),
\end{equation}
where $^{89}K({\rm main})$ is the Knight shift of the 
main $^{89}$Y spectral line measured by Mahajan {\it et al.}
\cite{Mahajan99} 
in Zn substituted 
YBa$_2$Cu$_3$O$_{6.6}$. 
The solid points in Fig.~8(b) indicate 
$^7K_0$ estimated this way. 
The dashed curve in Fig.~8(b) 
shows the estimate for $^7K_0$ obtained by using the 
$^{89}K$ measurements on pure 
YBa$_2$Cu$_3$O$_{6.63}$ \cite{Alloul89}.
In the Zn substituted samples, $^{89}K({\rm main})$ is 
slightly shifted with respect to $^{89}K$ of the pure samples. 
This might be due to the difference in the 
stoichiometry of the Zn substituted and the pure samples,
as discussed in Ref.~\cite{Mahajan99}. 

The results of the calculations on $\delta^7 K$ are represented 
by the various curves in 
Fig.~8(a).
The term subtracted from $^7K$ to obtain these results is 
shown by the solid line in Fig.~8(b).
Note that the solid line in Fig.~8(b) lies slightly 
above the solid line in Fig.~6(b), since 
larger values of $U$ were used
for the underdoped 
YBa$_2$Cu$_3$O$_{6+x}$.
These calculations are shown for four different 
values of $V_1$: $-0.125t$ (short-dashed), $-0.15t$ (solid),
$-0.175t$ (long-dashed), and 
$0.0$ (dot-dashed).
These results have been obtained by using $U(T)$ 
represented by the solid curve in Fig.~2(b).
Hence, 
they correspond to the first scenario described above where 
$T_2^{-1}$ has been linearly extrapolated for 
$T< 160$ K.
For each value of $V_1$, 
the calculations of $\delta ^7 K$ have been repeated 
according to the second scenario, where the saturation of 
$T_2^{-1}$ below 160 K is attributed to the saturation
of the antiferromagnetic correlations.
In this case, the values of $U(T)$ shown by the dotted curve 
in Fig.~2(b) were used.
In Fig.~8(a), these results are shown by the dotted curves
for each value of $V_1$ as indicated.
Comparing the results of the calculations with the experimental 
data, one sees that the 
values of $V_1$ between $-0.125t$ and $-0.15t$ would produce 
a fit of the experimental data
on $\delta^7K$.

Next, the results for $^{89}$Y are discussed.
The experiments by Mahajan {\it et al.} \cite{Mahajan94,Mahajan99}
found that Zn substitution in underdoped
YBa$_2$Cu$_3$O$_{6+x}$ strongly modifies the nuclear resonance 
spectrum of $^{89}$Y.  
In this case, in addition to the main $^{89}$Y resonance 
line, two satellite peaks are observed.
These outer and middle resonance lines have been 
identified as belonging to the $^{89}$Y 
nuclear spins which are respectively the first and the second 
near-neighbours of the Zn impurity,
Y(1) and Y(2) in Fig.~5.
The fact that the magnetic resonance spectrum of 
$^{89}$Y in Li substituted 
YBa$_2$Cu$_3$O$_{6+x}$
is nearly the same as in Zn substituted samples confirms 
this identification \cite{Bobroff}.
In the following, the data on $^{89}K$ 
will be compared with the results 
of the calculations which were outlined in Section III.
However, 
the comparisons will not be carried out directly for
$^{89}K$.
Rather, the shift of the satellite lines with respect to the 
main line will be used
in order to compensate for the 
opening of the pseudogap in the pure system.
Hence, the comparisons will be carried out for the following quantities,
\begin{eqnarray}
&& \delta^{89}K({\rm outer}) = ^{89}K({\rm outer})- ^{89}K({\rm main}) 
\nonumber \\ 
&& \delta^{89}K({\rm middle}) = ^{89}K({\rm middle})- ^{89}K({\rm main}). 
\end{eqnarray}
Since $^{89}K({\rm main})$ 
in Zn substituted material nearly
follows the $T$ dependence of 
$^{89}K$
in the pure system,
$\delta^{89}K({\rm outer})$ and 
$\delta^{89}K(\rm middle)$ 
represent the change in the Knight shifts 
of the nuclear spins at the Y(1) and Y(2) sites, respectively, 
due to the substitution of the Zn impurity.

In Figures 9(a) and (b), 
the points represent the experimental data on 
$-\delta^{89}K(\rm outer)$ and 
$-\delta^{89}K(\rm middle)$, respectively, for underdoped  
YBa$_2$Cu$_3$O$_{6+x}$ measured by Mahajan {\it et al.} 
\cite{Mahajan94,Mahajan99}.
In these figures, the data points are shown for two 
different orientations of the magnetic field ${\bf H}$.
It is estimated that the $^{89}$Y hyperfine coupling
has an anisotropy of about 15\%, 
which could be the cause of the anisotropy 
of the data seen in Figs.~9(a) and (b) \cite{Mahajan99,Takigawa93}.
In Fig.~9(a), the error bars shown for the lowest and the highest 
temperature measurements are representative of 
the error bars for the other data points.
In Fig.~9(b), the data points are shown up to 130 K, 
since at higher 
temperatures the middle satellite is not resolved.

The curves in Figs.~9(a) and (b) represent 
the results of the calculations for 
$-\delta^{89}K({\rm outer})$ and 
$-\delta^{89}K({\rm middle})$, respectively.
These calculations have been carried out for the same values 
of $V_1$ as in Fig.~8(a) for 
$\delta^7K$ \cite{89Y}.
For $T < 160$~K, these calculations were carried out 
in two ways by using $U(T)$ shown by the solid 
and the dotted curves in Fig.~2(b).
In Figs.~9(a) and (b), it is seen that for
$V_1$ between $-0.125t$ and $-0.15t$, 
the data can be fitted.
Finally, in Fig.~10 the $T$ dependence of 
$^{63}K_c({\bf r}_i)$ in underdoped 
YBa$_2$Cu$_3$O$_{6+x}$ 
is shown for various Cu(2) sites near the impurity.
Here, $V_1=-0.15t$ was used. 
In order to interpret these,
in the next section the real-space structure of the changes 
induced in the magnetic correlations 
by the impurity will be shown.
The $^{63}K_c$ measurements in Zn substituted 
YBa$_2$Cu$_3$O$_{6.7}$ 
imply that the polarisation around the impurity 
is staggered \cite{Julien}. 

\subsection{Pattern of the magnetic correlations 
around the impurity}

In Fig.~11(a), $k({\bf r})$ defined by 
Eq.~(\ref{kr}), is shown as a function of the lattice distance 
$r=|{\bf r}|$ away from the impurity at $T=100$ K.
Here, $r$ is given in units of the lattice constant, and 
in obtaining these results $V_1=-0.15t$ 
was used.
The horizontal long-dashed line denotes what is expected for 
$k({\bf r})$ within this model for pure 
optimally doped
YBa$_2$Cu$_3$O$_{6+x}$,
and the open circles represent $k({\bf r})$ 
when an impurity is introduced into this material 
at ${\bf r}_0=0$.
The results for underdoped 
YBa$_2$Cu$_3$O$_{6+x}$
are also shown in Fig.~11(a).
The dotted horizontal line denotes 
$k({\bf r})$ for pure underdoped 
YBa$_2$Cu$_3$O$_{6+x}$, 
while the filled circles show 
$k({\bf r})$ within the presence of the impurity
for this material.
The results on underdoped 
YBa$_2$Cu$_3$O$_{6+x}$
were obtained using the value of $U$ 
given by the solid curve in Fig.~2(b).
Note that, for underdoped 
YBa$_2$Cu$_3$O$_{6+x}$ in this graph, 
the physically more relevant quantity is the change in 
$k({\bf r})$ induced by the impurity, since the presence of the 
pseuogap in the uniform susceptibility 
of the pure system is not included in the model.

The real-space structure of $k({\bf r})$ is important since 
it determines the Knight shift at various sites.
For instance, 
$k({\bf r})$ at ${\bf r}=(1,0)$ determines the value of 
$^7K$.
On the other hand, 
$\delta^{89}K({\rm outer})$, corresponding to the 
Y(1) site in Fig.~5, is given by the change in 
\begin{equation}
k(0,0) + 2k(1,0) + k(1,1)
\end{equation}
induced by the impurity.
Similarly, 
$\delta^{89}K({\rm middle})$, corresponding to the Y(2) 
site, probes the change induced in 
\begin{equation}
k(1,0) + k(1,1) + k(2,0) + k(2,1)
\end{equation}
by the impurity.
Hence, it is seen that by the fitting of 
$\delta ^7K$ in Fig.~8(a), the value of 
$k(1,0)$ is fixed.
Since it is known that $k(0,0)$ vanishes when the impurity is 
introduced, the fitting of 
$\delta^{89}K({\rm outer})$
in Fig.~9(a) fixes the value of $k(1,1)$.
In Fig.~11(a), it is seen that while 
$k(1,0)$ is enhanced strongly 
by the impurity, 
$k(1,1)$ is strongly suppressed and it even becomes negative.
The fact that both $\delta^7 K$ and 
$\delta^{89}K({\rm outer})$ can be fitted 
at the same time means that the induced changes in 
$k({\bf r})$ of the actual system must have a staggered pattern at 
the sites (1,0) and (1,1).
Since $k(1,0)$ and $k(1,1)$ are now fixed, 
the fact that 
$\delta^{89}K({\rm middle})$
can also be fitted means that the change in 
$k(2,0)+k(2,1)$ is fixed.
In Fig.~11(a), it is seen that $k(2,0)$ gets suppressed
by the impurity and $k(2,1)$ gets enhanced while the total 
effect on $k(2,0)+k(2,1)$ is small.
This in turn means that the induced changes in 
$k({\bf r})$ of the actual system must also be staggered 
at sites (2,0) and (2,1).
Hence, the fits obtained in Figs.~8(a), 9(a) and 
(b) imply that $k({\bf r})$ has a staggered 
pattern in the vicinity of the impurity.
This conclusion was reached by the 
analysis of Mahajan {\it et al.} 
\cite{Mahajan94,Mahajan99}.
Here it is seen that this simple model produces 
this pattern as well.

It is useful to compare the staggered pattern of $k({\bf r})$ 
with the pattern of $\chi({\bf r},{\bf r})$ near the impurity.
Fig.~11(b) shows $\chi({\bf r},{\bf r})$ 
versus $r$ plotted in the same way as $k({\bf r})$ 
is plotted in Fig.~11(a).
Here one sees that $\chi({\bf r},{\bf r})$ 
is strongly modified near the impurity.
It is also seen that $\chi({\bf r},{\bf r})$ 
does not have a staggered pattern as, for instance, 
one would have had for antiferromagnetically ordered spins.
The comparisons shown in Figs.~6, 8 and 9 mean that 
the experimental data on $^7K$ and $^{89}K$ require 
a staggered pattern for $k({\bf r})$
in the vicinity of the impurity.
But a staggered structure for 
$\chi({\bf r},{\bf r})$ is not necessary.

In Figs.~11(a) and (b), 
it is seen that 
while $\chi({\bf r},{\bf r})$ is positive for all 
${\bf r}$, $k({\bf r})$ can become negative, for instance at 
${\bf r}=(1,1)$.
This is possible because $\chi({\bf r},{\bf r'})$ for 
${\bf r}=(1,0)$ and ${\bf r'}=(1,1)$ has a large 
negative value after the impurity is introduced, 
meaning that the impurity is inducing strong 
antiferromagnetic correlations in its local 
environment in this model.
These points will be discussed further in Section IV.A.

At this point it is also useful to present results on how the impurity 
potential affects the single-particle properties. 
For this purpose, in Fig.~11(c) the electron occupation number 
$n({\bf r}_i)= \sum_{\sigma} 
\langle c^{\dagger}_{i\sigma} c_{i\sigma} \rangle$
for sites near the impurity is shown as a function of 
the distance $r=|{\bf r}_i|$ from the impurity
for various values of $V_1/t$.
Since $V_0$ is strongly attractive, the impurity site
which is not shown in this figure, 
is nearly doubly occupied, $n({\bf r}_i=0)\approx 2.0$.
For $V_1=0$, it is seen that $n({\bf r}_i)$ is 
enhanced at ${\bf r}_i=(1,1)$ 
with respect to its value at ${\bf r}_i=(1,0)$.
Away from the impurity, these oscillation in $n({\bf r}_i)$
decay, and $n({\bf r}_i)$ goes to its value in the pure system,
$\langle n\rangle=0.86$,
which is indicated by the horizontal long-dashed line.
Here, it is also seen that this structure in $n({\bf r}_i)$ 
holds for weak attractive values of $V_1$.
But as $V_1$ becomes more attractive, for instance for 
$V_1=-0.15t$, the pattern in $n({\bf r}_i)$ changes. 
In this case, $n({\bf r}_i=(1,0))$ gets enhanced 
over $n({\bf r}_i=(1,1))$. 
Hence, in Fig.~11(c) one observes that the structure 
in $n({\bf r}_i)$ in the vicinity of the impurity
does not necessarily reflect the sign of $V_1$.
For instance, a finite attractive $V_1$ is required before 
$n({\bf r}_i=(1,0))$ gets enhanced 
with respect to $n({\bf r}_i=(1,1))$.
This has to do with the 
presence of the strongly attractive $V_0$ term 
in the impurity potential. 

In Section III.B, it was seen that the data 
on $\delta^7K$ and $\delta^{89}K$ can 
be fitted reasonably well over the whole 
temperature range by using a simple form 
for the effective impurity potential
without any $T$ dependence.
On the other hand, because of the various approximations employed
in this model, the quantitative agreement obtained 
in the fits might actually be misleading.
However, what is significant is the fact that 
$\delta^7K$, $\delta^{89}K({\rm outer})$ and 
$\delta^{89}K({\rm middle})$ can be fitted all
at the same time, which means that the real-space
structure of the changes induced in the magnetic correlations
near the impurity 
appears to be described by this simple model.

\section{Discussion}

\subsection{Role of the antiferromagnetic correlations}

It is useful to discuss the connection between the antiferromagnetic
correlations of the system 
which is composed of the impurity and the host,
and the anomalous enhancement 
of the Knight shift at sites around the impurity.
This connection becomes more clear if the Fourier transform 
$\chi({\bf q},{\bf q'})$ is introduced,
\begin{equation}
\label{chirr}
\chi({\bf q},{\bf q'})=\sum_{{\bf r},{\bf r'}} 
e^{ i({\bf q}\cdot{\bf r} - {\bf q'}\cdot {\bf r'}) }
\chi({\bf r},{\bf r'}).
\end{equation}
Note that $\chi({\bf r},{\bf r'})$ 
is the susceptibility 
for an $N$-site square lattice with one impurity 
at the center and periodic boundary conditions.

An especially important quantity is 
the off-diagonal susceptibility 
$\chi({\bf q},0)\equiv
\chi({\bf q},{\bf q'}=0)$,
since its Fourier transform with respect to 
${\bf q}$ gives $k({\bf r})$.
Particularly, 
at sites ${\bf r}=(1,0)$ and $(1,1)$
with respect to the impurity,
one has 
\begin{eqnarray}
\label{k10}
k(1,0) && = \sum_{\bf q} {1\over 2} ( \cos{q_x} + \cos{q_y} )
\, \chi({\bf q},0) \nonumber \\
k(1,1) && = \sum_{\bf q} \cos({q_x+q_y}) 
\, \chi({\bf q},0).
\end{eqnarray}
In Figs.~12(a) and (b), 
the ${\bf q}$ dependence of 
$-\chi_0({\bf q},0)$, which is for $U=0$, and
$-\chi({\bf q},0)$ are shown at 100~K for 
optimally doped 
YBa$_2$Cu$_3$O$_{6+x}$
obtained by using $V_1=-0.15t$ \cite{Finitesize}.
Here, 
the $\delta$-function component of 
$\chi({\bf q},{\bf q'}=0)$ at ${\bf q}=0$ has been 
omitted.
In these figures, it is seen that 
$-\chi({\bf q},0)$  is enhanced with respect to 
$-\chi_0({\bf q},0)$, and they both peak at 
${\bf q}$ near $(\pi,\pi)$.
Note that for ${\bf q}\sim (\pi,\pi)$, the form 
factors entering Eq.~(\ref{k10}), 
${1\over 2} ( \cos{q_x} + \cos{q_y} ) \sim -1$ 
and 
$\cos(q_x+q_y)\sim 1$.
Because of the ${\bf q}$ structure of $\chi({\bf q},0)$, 
$k(1,0)$ gets enhanced and $k(1,1)$ gets suppressed.
Hence, within this model, 
the staggered pattern of $k({\bf r})$ near the impurity
is a consequence of the 
peaking of $-\chi({\bf q},0)$ near $(\pi,\pi)$.

The enhancement of $-\chi({\bf q},0)$ with respect to
$-\chi_0({\bf q},0)$ is understood better if 
the RPA equation, Eq.~(\ref{rpa}), 
is written in momentum space,
\begin{equation}
\label{chiqq}
\chi({\bf q},{\bf q'}) = 
\chi_0({\bf q},{\bf q'})  + 
U \sum_{{\bf q''}} 
\chi_0({\bf q},{\bf q''}) 
\chi({\bf q''},{\bf q'}),
\end{equation}
where 
$\chi_0({\bf q},{\bf q'})$ is for $U=0$.
For the pure system, 
$\chi({\bf q},{\bf q'}=0)$ is given by 
\begin{equation}
\chi({\bf q},0) 
= N \delta_{{\bf q},0} \chi_{pure} ( {\bf q}\rightarrow 0)
\end{equation}
where $\chi_{pure}$ is defined by Eq.~(\ref{chirpa}),
and within the presence of the impurity
$\chi({\bf q},{\bf q'}=0)$ is obtained by solving 
\begin{equation}
\label{chiq0}
\chi({\bf q},0) = 
\chi_0({\bf q},0)  + 
U \sum_{{\bf q''}} 
\chi_0({\bf q''},0) 
\chi({\bf q''},{\bf q}). 
\end{equation}
This expression shows that when ${\bf q''}\sim (\pi,\pi)$, 
$\chi({\bf q}\sim(\pi,\pi),{\bf q'}=0)$ couples 
to the antiferromagnetic correlations
determined by 
$\chi({\bf q},{\bf q})$ with ${\bf q}\sim (\pi,\pi)$
of the system which is composed of the host and the impurity.
This is the reason for the strong enhancement of
$\chi({\bf q},{\bf q'}=0)$ with respect to 
$\chi_0({\bf q},{\bf q'}=0)$.

It is useful to discuss the physical meaning of the off-diagonal 
susceptibility $\chi_0({\bf q},{\bf q'})$ where
${\bf q}\neq{\bf q'}$.
During the scattering of the spin fluctuations by the impurity 
potential, the momentum is not conserved and 
$\chi_0({\bf q},{\bf q}\neq {\bf q'})$ acts as the 
vertex for the scattering of the spin fluctuations 
by the impurity
with ${\bf Q^*}={\bf q}-{\bf q'}$ momentum transfers.
The peaking of $\chi_0({\bf q},{\bf q'}=0)$
near ${\bf q}\sim (\pi,\pi)$ means that the scattering 
of the antiferromagnetic spin fluctuations with momentum transfers 
near $(\pi,\pi)$ is the dominant scattering process.
The effects of the scatterings with large momentum transfers 
on the ${\bf Q}=(\pi,\pi)$ neutron scattering intensity have been
also emphasized in Ref. \cite{neutron},
where $\chi_0({\bf q},{\bf q'})$ was calculated at the lowest
order in the impurity potential.
Within this model, both the Knight shifts and the neutron 
scattering experiments point out at the importance of the 
scattering of the particle-hole pairs by the nonmagnetic 
impurity with large momentum transfers near $2{\bf k}_F$.
If the $\sim 2{\bf k}_F$ scatterings are indeed one of the primary 
effects of the nonmagnetic impurities, 
then an anomalous softening of the phonons at 
wave vectors $\sim 2{\bf k}_F$ might be observed in 
YBa$_2$Cu$_3$O$_{6+x}$ with dilute Zn
impurities \cite{SidisPC}.

The real-space
pattern of $^{63}K_c({\bf r}_i)$ seen in Fig.~7(b)
also reflects the ${\bf q}$ dependence of 
$\chi({\bf q},{\bf q'}=0)$.
Equation~(\ref{Kc}) for $^{63}K_c({\bf r}_i)$
can be rewritten as 
\begin{equation}
^{63}K_c({\bf r}_i)= 
{1\over 2}
\bigg( {\gamma_e \over ^{63}\gamma_n } \bigg)
{1 \over N}
\sum_{\bf q} \,
e^{ i{\bf q} \cdot {\bf r}_i }
\big(
A_c + 4B\gamma_{\bf q} 
\big)
\chi({\bf q},0),
\end{equation}
where $\gamma_{\bf q} = (\cos{q_x} + \cos{q_y})/2$.
Since $\chi({\bf q},{\bf q'}=0)$ has most of its weight at 
${\bf q}\sim (\pi,\pi)$ for which
$\gamma_{\bf q}\sim -1$,
one gets 
\begin{equation}
^{63}K_c({\bf r}_i) \approx -
{1\over 2}
\big(
|A_c| + 4B 
\big)
\bigg( {\gamma_e \over ^{63}\gamma_n } \bigg)
{1 \over N}
\sum_{{\bf q} \sim (\pi,\pi)} \,
e^{ i{\bf q} \cdot {\bf r}_i }
\chi({\bf q},0).
\end{equation}
Here one sees 
that $|A_c|+4B$ has a large value and 
$^{63}K_c({\bf r}_i)$ directly couples to 
$\chi({\bf q}\sim(\pi,\pi),{\bf q'}=0)$.
It has been noted in Ref.~\cite{Julien} that the 
anomalous broadening of the $^{63}$Cu(2) linewidth 
reflects the staggered 
polarization of the magnetic correlations near the impurity. 

It is important to note that, within this model, 
the antiferromagnetic correlations also get enhanced due to the 
presence of the impurity \cite{Bulut2000}.
If $N_i$ randomly distributed nonmagnetic impurities 
are considered in the dilute limit, 
then the ${\bf q}$-dependent magnetic susceptibility 
of this system is given by 
\begin{equation}
\chi({\bf q}) \equiv
\chi_{pure}({\bf q}) + 
n_i N 
\bigg[ 
\chi({\bf q},{\bf q}) - \chi_{pure}({\bf q}) 
\bigg],
\end{equation}
where $n_i=N_i/N$ is the impurity concentration.
The results on $\chi({\bf Q^*})$ for 
0.5\% nonmagnetic impurities are compared with 
$\chi_{pure}({\bf Q^*})$
in Figs.~3(b) and (c),
where it is seen that the impure system has stronger 
antiferromagnetic correlations than the pure system.
Here, ${\bf Q^*}$ is the incommensurate wave vector where 
$\chi$ peaks at low temperatures.
The enhancement of $\chi({\bf Q^*})$ over 
$\chi_{pure}({\bf Q^*})$ is especially significant 
for the underdoped system.
Hence, it needs to be noted that the enhancement of the Knight 
shifts near the impurity in this model 
is a result of the 
coupling to $\chi({\bf q},{\bf q})$ for ${\bf q}\sim (\pi,\pi)$
of the whole system
which is composed of the impurity and the host 
rather than to $\chi_{pure}({\bf q}\sim(\pi,\pi))$ 
of the pure host.

Experimentally, 
${\rm Im}\,\chi({\bf Q},\omega)$ is the inelastic 
neutron scattering spectral weight. 
The enhancement of 
${\rm Im}\,\chi({\bf Q},\omega)$ 
in the normal state by the impurity scattering is clearly 
seen for dilute Zn impurities in
YBa$_2$Cu$_3$O$_{7}$
\cite{Sidis,Fong,Sidis2000}.
The low frequency part of ${\rm Im}\,\chi$ is also probed by the 
measurement of the longitudinal relaxation rate 
$T_1^{-1}$ at sites near the impurity.
The measurements of $T_1^{-1}$ at sites near the 
impurity have found interesting results
\cite{Mahajan94,Mahajan99,MacFarlane,Julien}.
The calculations presented here 
have been extended to obtain the $^7$Li $T_1^{-1}$
using the Pade approximation
for analytic continuation to the real-frequency 
axis of the results calculated in terms of the 
Matsubara frequencies.
While it is difficult to obtain reliable results on 
${\rm Im}\,\chi({\bf Q},\omega)$ for general $\omega$ 
using the Pade approximation,
it is possible to obtain control on the calculation
of $T_1^{-1}$, which requires only 
the $\omega\rightarrow 0$ limit.
Since the analytic continuation procedure requires special
attention, the calculations of 
the $^7$Li $T_1^{-1}$ 
will be presented elsewhere \cite{Korringa,Li7}.

\subsection{Effects of the pseudogap in underdoped 
YBa$_2$Cu$_3$O$_{6+x}$}

The analysis of the data on underdoped YBa$_2$Cu$_3$O$_{6+x}$
needs to be interpreted carefully because of the presence 
of the pseudogap in this system.
The uniform susceptibility of 
pure YBa$_2$Cu$_3$O$_{6.63}$ 
starts to decrease below 300 K, with a $T$ dependence 
proportional to that given
by the dashed curve in Fig.~8(b).
This has been ignored in the calculations presented above.
In order to explore the effects of the pseudogap on the Knight 
shift results, the following simple 
calculation has been carried out.
The suppression of the diagonal irreducible susceptibility 
$\chi_0({\bf q},{\bf q})$ for ${\bf q}\sim 0$ 
has been artificially incorporated 
into the model by multiplying 
$\chi_0({\bf q},{\bf q})$ by the factor
\begin{equation}
\label{F}
F({\bf q}) = 
1 - a e^{-|{\bf q}|^2/\kappa^2},
\end{equation}
which has been chosen in order to cause a suppression 
for ${\bf q}\sim 0$.
After artificially suppressing the diagonal 
irreducible susceptibility for ${\bf q}\sim 0$, 
the Fourier transform is taken to obtain 
$\chi_0({\bf r},{\bf r'})$ of the pure system, which 
is then used in calculating $\chi_0({\bf r},{\bf r'})$ 
of the impure system.
Hence, in this procedure, the diagonal terms 
$\chi_0({\bf q},{\bf q})$ for ${\bf q}\sim 0$ 
of the impure system have the effects of the pseudogap, 
but $\chi_0({\bf q},{\bf q})$ for ${\bf q}\sim (\pi,\pi)$ 
and the off-diagonal terms 
$\chi_0({\bf q},{\bf q'})$, where ${\bf q}\neq {\bf q'}$, 
are not affected by this artificial opening of the pseudogap.
Next, $\chi_0({\bf r},{\bf r'})$ calculated with this 
procedure was used in solving Eq.~(\ref{rpa}) 
for $\chi({\bf r},{\bf r'})$
and in determining the Knight shifts.
The parameter $\kappa$ entering Eq.~(\ref{F})
was arbitrarily chosen to be $\pi/2$.
In addition, $a$ has been chosen such that the ratio of 
\begin{equation}
{ F(0) \chi_0^L(0) 
\over 
1 - U F(0) \chi_0^L(0) }
\end{equation}
to $\chi_0^L(0)/(1-U\chi_0^L(0))$ is equal to 
the ratio of $^{89}K$ in pure underdoped 
YBa$_2$Cu$_3$O$_{6+x}$
to that in pure optimally doped 
YBa$_2$Cu$_3$O$_{6+x}$.
For instance, at $T=100$ K, this condition is satisfied 
by using $a=0.51$.
The opening of the pseudogap 
in this way requires slightly larger values of $V_1$ 
for fitting the Knight shift data. 
In Section III.B, the data on $\delta^7K$, 
$\delta^{89}K({\rm outer})$ and 
$\delta^{89}K({\rm middle})$ were fitted by using $V_1$ between 
$-0.125t$ and $-0.15t$.
Here, if $V_1=-0.15t$ is used along with the above 
mentioned values of $\kappa$ and $a$, then at $T=100$~K one 
obtains the following results for the Knight shifts:
$\delta^7K=650$ ppm, 
$\delta^{89}K({\rm outer})=-180$ ppm, and 
$\delta^{89}K({\rm middle})=-75$ ppm.
On the other hand, at 300 K, it is necessary to use $a=0.25$ 
and in this case one obtains:
$\delta^7K=250$ ppm, 
$\delta^{89}K({\rm outer})=-35$ ppm, and 
$\delta^{89}K({\rm middle})=-20$ ppm.
These values for the Knight shifts 
are comparable to the data seen in 
Figs.~8(a) and 9.
Hence, when the pseudogap is introduced 
in $\chi_0({\bf q},{\bf q})$ for ${\bf q}\sim 0$ 
in this artificial way, the Knight shift data can be still fitted, 
but by using a slightly larger value for $V_1$.
However, 
it must be kept in mind that 
the way the pseudogap is introduced here is not rigorous, 
and in fact the fitting of the data in this way can be
considered as superfluous. 
For this reason, the fitting of the data on the underdoped 
compound will not be pursued further. 
Rather, here it is only pointed out that
the staggered nature of the induced 
magnetic correlations seen in this model appears to be consistent
with the measurements of the 
$^7$Li and $^{89}$Y Knight shifts in underdoped 
YBa$_2$Cu$_3$O$_{6+x}$.

\subsection{Dependence of the results 
on the effective bandwidth}

The results seen in Section III were obtained by using an effective
bandwith $W$ of 1 eV. 
For a one-band model of the cuprates, 
the bare hopping matrix element is estimated
to be of order 0.45 eV leading to a bare bandwidth of 3.6 eV.
While in principle one expects that the Coulomb correlations 
act to reduce the bandwidth, it is not clear what the precise 
value of the effective bandwidth should be
in an RPA framework.
Hence, it is necessary to check the dependence
of the results on the effective bandwidth $W$, 
which was assumed to be 1 eV in Section III.
For this reason, here, the Knight shifts are calculated 
for $W=3$~eV. 
In this case, larger values of $U$ are required for fitting the 
$T_2^{-1}$ data and, consequently, the pure system has 
a stronger Stoner enhancement of the antiferromagnetic correlations. 
For $W=3$~eV, the value of 
the $^{89}$Y hyperfine coupling $D$ was chosen to be
$-2$ kOe/$\mu_B$ in order to fit the value of $^{89}K$ for pure 
YBa$_2$Cu$_3$O$_{7}$
in the normal state. 
The experimental estimate for $D$ is also about 
$-2$ kOe/$\mu_B$ \cite{Mahajan99}.

In Fig.~13, $^7K$ versus $T$ is shown for optimally doped 
YBa$_2$Cu$_3$O$_{6+x}$ calculated using $W=3$~eV.
Here, the $^7$Li hyperfine coupling $C$ 
was taken to be $1.4$ kOe/$\mu_B$.
The open circles represent $^7K_0$ deduced from the 
$^{89}K$ data \cite{Alloul89} on pure 
YBa$_2$Cu$_3$O$_{7}$.
The dotted curve represents the theoretical results for 
$^7K_0$. 
Here, it is seen that a smaller value of 
$V_1$ is required for fitting the $^7K$ data as compared to
that for $W=1$ eV.
This is because in this case the antiferromagnetic
correlations are stronger. 

It is also necessary to compare the value of $C$ used here 
with the experimental estimate.
In Ref.~\cite{Bobroff}, $C$ was estimated to be 2.4 kOe/$\mu_B$.
This value was deduced by comparing the enhancement of the 
uniform susceptibility $\Delta\chi$ with the enhancement of 
$^7K$ for the underdoped 
YBa$_2$Cu$_3$O$_{6+x}$
by assuming that the contribution to $\Delta\chi$ is 
arising from the changes in the magnetic correlations 
at only the four nearest-neighbour Cu(2) sites 
of the impurity. 
Note, however, that $\Delta\chi$ is given by 
\begin{equation}
\Delta\chi = \sum_{j} k({\bf r}_j) - 
\chi_{pure}({\bf q}\rightarrow 0),
\end{equation} 
where the sum over $j$ is carried over the whole lattice, 
and the Knight shift data and the numerical results presented 
in this paper indicate that the changes induced in $k({\bf r}_j)$ 
by the impurity is extended and not restricted to the four 
nearest-neighbor sites of the impurity.
In fact, by limiting the changes in $k({\bf r}_j)$ to be only at the 
nearest-neighbor sites, one would overestimate 
$\Delta\chi$. 
Similarly, when one calculates $C$ 
by using the experimental data on 
$\Delta\chi$ and the $^7K$ data, 
one would overestimate $C$.
For instance, in Fig. 11(a), it is seen that at 100~K
the enhancement of $k(1,0)$ by the impurity, 
$\Delta k(1,0)$, is about 20 states/eV
for underdoped 
YBa$_2$Cu$_3$O$_{6+x}$.
If $\Delta\chi$ is estimated by just using the change in 
$k(1,0)$, then one would obtain $\Delta\chi=0.4$ states/eV
in the dilute limit for an impurity concentration of 0.5\%.
On the other hand, the calculated value of $\Delta\chi$ at 
$T=100$ K and $V_1=-0.15t$ for underdoped 
YBa$_2$Cu$_3$O$_{6+x}$ in this model is 
$\sim 0.2$ states/eV.
Hence, by using $\Delta k(1,0)$ 
and assuming that the changes in $k({\bf r})$ occur 
only at the nearest-neighbor sites of the impurity, one would 
overestimate $\Delta\chi$ by about a factor of two.
Similarly, 
if $\Delta k(1,0)$ and $\Delta\chi$ are used 
to calculate $C$ one would overestimate $C$ by the same amount.
When this is taken into account, it is seen that the values of 
$C$ used here, 0.85 kOe/$\mu_B$ in Section III and 
1.4 kOe/$\mu_B$ in Section IV.C, are comparable to the experimental 
estimate which was obtained by using 
the data on $\Delta\chi$ and $^7K$.

In Figs.~14(a)--(c), the $T$ dependence of 
$\delta^7K$, $-\delta^{89}K$(outer), and 
$-\delta^{89}K$(middle) are shown for underdoped 
YBa$_2$Cu$_3$O$_{6+x}$.
Here, it is seen that for $V_1=-0.03t$, 
the magnitudes of the calculated Knight shifts 
agree with the experimental data.
However, the fitting of the $T$ dependence 
of the Knight shifts 
is not as good as that seen in Section III for $W=1$ eV.

The results presented in this section show that the value of 
$V_1$ required for fitting the Knight shift data depends 
on the effective bandwidth. 
For $W=3$~eV, the system has stronger enhancement of the AF
correlations compared to that for $W=1$~eV.
Here, it has been shown that as $W$ increases $V_1$ required 
for fitting the data decreases.
However, even if a $W$ of 4~eV is used, at the level of RPA, 
an impurity potential with $V_1=0$ does not produce sufficient 
$T$ variation for fitting the $^7K$ data on optimally doped
YBa$_2$Cu$_3$O$_{6+x}$,
and a weakly attractive $V_1$ is necessary.

\subsection{Nature of the effective impurity potential}

Because of the Coulomb correlations, 
the bare impurity potential which has only an onsite 
component $V_0\sum_{\sigma} c^{\dagger}_{0\sigma}c _{0\sigma}$
could acquire an extended component
through higher-order scattering processes \cite{Ziegler}.
How this could happen at lowest order in the Coulomb 
repulsion $U$ is discussed in Appendix II.
This extended component is modelled here by using a 
static $V_1$ as a free parameter.
In this picture, the fact that the Knight shift 
data imply $V_1<0$ is giving information on the 
effective particle-particle interaction in the system.
The real-space structure of the effective impurity interaction 
was studied at half-filling for the $t$-$J$ model 
within the exact diagonalization calculations \cite{Ziegler}.
It is useful to calculate $V_1$ 
in exact numerical calculations away from half-filling.
This would be a test of the model presented here.
Furthermore, note that here $V_1$ is assumed to have no 
temperature dependence.
However, $V_1$ could depend on $T$,
if indeed the Coulomb correlations play 
a role in inducing the extended component of the impurity potential.
This could quantitatively affect the fits seen above.

These results carried out at the level of RPA imply that 
$V_1$ is weakly attractive.
At this level, even though an onsite scattering potential 
($V_1=0$) yields large enhancements of the Knight shifts, 
it is still insufficient for fitting the $T$ dependence,
for instance, of $^7K$ in optimally doped 
YBa$_2$Cu$_3$O$_{6+x}$.
However, note that corrections beyond RPA 
might change this result. 
For instance, the spin-fluctuation self-energy corrections
to the single-particle Green's functions, 
when taken into account self-consistently, 
could play an important role.

Within this model, the Knight shift experiments on the cuprates with 
nonmagnetic impurities are interesting especially 
because they probe the interplay of the correlations in the 
density and the magnetic channels.
The effective impurity potential acts in the density channel while
its spatially-resolved magnetic response is detected  
through the Knight shift measurements.
Similarly, the recent $^7$Li Knight shift 
measurements \cite{NMRSC}
in the $d$-wave superconducting state of 
YBa$_2$Cu$_3$O$_{6+x}$
give valuable information  
about the interplay of the density, magnetic and the $d$-wave 
superconducting correlations in this material.

Another set of experiments which produce spatially 
resolved information on the effects of the nonmagnetic impurities 
are the STM measurements on Zn substituted BISCO
\cite{Pan}.
These measurements have been carried out in the superconducting 
state. 
Clearly, similar STM measurements above $T_c$ would be 
useful for understanding the effects of the nonmagnetic impurities
in the normal state.
This would allow for a direct comparison of the single-particle
properties with this model.

It is also desirable to extend these model calculations 
to the superconducting state in order to make comparisons 
with the STM \cite{Pan} and the recent NMR \cite{NMRSC}
measurements below $T_c$.
However, note that if a spin-fluctuation mediated mechanism 
is assumed for the $d$-wave pairing, then the effects of the impurity 
on the pairing potential 
need to be taken into account as well as the scattering 
of the quasiparticles by the impurity.
This is because it is already known in the normal state 
that in the local environment of the impurity the spin-fluctuations 
are strongly modified 
\cite{Mahajan94,Bobroff,Mendels,Mahajan99,Alloul99,MacFarlane,Julien,Ishida93,Zheng,Ishida96}.

Here, the effects of the scatterings by a nonmagnetic impurity 
on the magnetic spectrum is studied by modelling the impurity 
as a potential scatterer. 
In this respect, disorder could have effects similar to those of 
nonmagnetic impurities, and it has been already pointed out
that the anomalous 
line broadening of $^{63}$Cu(2) in 
La$_{2-x}$Sr$_x$CuO$_4$ could be due 
to the intrinsic disorder in this compound 
\cite{Slichter}.

\section{Summary and Conclusions}

In this paper, the Knight shift data on
YBa$_2$Cu$_3$O$_{6+x}$
with nonmagnetic impurities 
have been analyzed within a rather simple model
exhibiting short-range antiferromagnetic correlations.
The antiferromagnetic correlations have been modelled
within the framework of the 2D Hubbard model, 
and the effects of an impurity have been approximated by
using a static extended impurity potential.
The strength of the antiferromagnetic correlations in the 
pure system has been determined by fitting the 
$T_2^{-1}$ data of pure 
YBa$_2$Cu$_3$O$_{6+x}$.
The impurity potential has been assumed to have a range of 
just one lattice spacing without any temperature dependence.
The onsite component of the impurity potential was taken to be 
strongly attractive, and the near-neighbor component $V_1$
was treated as a free parameter 
in fitting the Knight shift data. 
The simplicity of the model and the differences 
in the stoichiometry of the samples along 
with the uncertainities in the hyperfine couplings 
are factors which limit the conclusions which can be drawn.
Nevertheless, here,
it has been found that the anomalous $T$ dependence of $^7 K$ in 
YBa$_2$Cu$_3$O$_{7}$
can be fitted, 
if a weakly attractive $V_1$ is used. 
The nature of the effective impurity potential 
could be studied in the 
paramagnetic state of the Hubbard model, where there are
short range antiferromagnetic correlations,
by using exact numerical methods.
This could be a test of one of the main assumptions 
of this model.
Especially, the sign of $V_1$ could be tested.
These calculations have been 
also extended to the case of underdoped 
YBa$_2$Cu$_3$O$_{6+x}$
after making various assumptions about the magnetic correlations 
in this compound, which has a magnetic pseudogap.
Based on these assumptions, 
it has been found that 
the real-space structure of the magnetic correlations
in the vicinity of the impurity is consistent with the
Knight shift experiments.
However, caution is necessary in interpreting the 
results on underdoped 
YBa$_2$Cu$_3$O$_{6+x}$.

The results presented in this paper depend on the nature of the 
effective impurity potential used.
Any magnetic scattering component in $V_{\rm eff}$
could significantly change the results.
Furthermore, it needs to be kept in mind that 
the validity of these results depend on the 
weak-coupling approach used for calculating the magnetic correlations.
Clearly, much remains to be understood about the 
effects of the Zn or Li impurities on the magnetic correlations of 
YBa$_2$Cu$_3$O$_{6+x}$
in the normal state.

\vspace{0.4 in}

\begin{center}
{\bf {Appendix I: {\boldmath$^7$}Li Knight shift}}
\end{center}

In the presence of a uniform static magnetic field, 
${\bf B}=B_0{\bf z}$, and a hyperfine coupling to the 
electronic spins given by Eq.~(\ref{A7}),
the interaction energy of a $^7$Li nuclear moment 
$^7{\bf {\mu}}= \hbar ^7\gamma_n {\bf I}$ is 
\begin{equation}
-\hbar ^7\gamma_n I^z B_0 + 
C I^z \sum_{i=1}^4 
\langle S^z({\bf r}_i) \rangle, 
\end{equation}
where $i$ sums over the four Cu(2) sites neighboring the $^7$Li 
impurity. 
This expression can be rewritten as 
\begin{equation}
-\hbar ^7\gamma_n I^z B_0 ( 1 + ^7K ),
\end{equation}
where the $^7$Li Knight shift $^7K$ defined by 
\begin{equation}
^7K = - 
\bigg( { C \over \hbar ^7\gamma_n B_0 } \bigg)
\sum_{i=1}^4 \langle S^z({\bf r}_i) \rangle 
\end{equation}
gives the fractional change in the Zeeman frequency 
of the nuclear magnetic moment due to the hyperfine coupling. 
According to the Kubo linear-response theory \cite{Mahan}, 
the expectation value at time $t$
of the electronic spin at site ${\bf r}_i$,
$\langle S^z({\bf r}_i,t)\rangle $,
is given by 
\begin{equation}
\label{Sz}
\langle S^z({\bf r}_i,t) \rangle = -i
\int_{-\infty}^t \, dt' \,
\langle [ S^z({\bf r}_i,t), V_Z(t')] \rangle 
\theta(t-t') ,
\end{equation}
where $V_Z(t') = e^{iH_0 t} V_Z e^{-iH_0 t}$ with 
the Zeeman term for the electronic spins
\begin{equation}
V_Z = - \hbar \gamma_e B_0 \sum_j S^z({\bf r}_j).
\end{equation}
In Eq.~(\ref{Sz}), the expectation values are evaluated with 
respect to the eigenstates of
\begin{equation}
H_0 = H_{Hubbard} + V_{imp},
\end{equation}
where $H_{Hubbard}$ is given by Eq.~(\ref{Hubbard}) and the impurity 
interaction $V_{imp}$ is given by Eq.~(\ref{Vimp}).
In the adiabatic limit and as $t\rightarrow \infty$, 
$\langle S^z({\bf r}_i) \rangle $
is related to the magnetic susceptibility 
\cite{Mahan},
and in this limit 
Eq.~(\ref{Sz}) reduces to 
\begin{equation}
\langle S^z ({\bf r}_i) \rangle = 
{1\over 2} \hbar \gamma_e B_0 
\sum_j \chi({\bf r}_i, {\bf r}_j)
\end{equation}
where the transverse susceptibility 
$\chi({\bf r}_i,{\bf r}_j)= \chi({\bf r}_i,{\bf r}_j,i\omega_m=0)$
is defined by Eq.~(\ref{chi}).
Hence, the $^7$Li Knight shift is given by 
\begin{equation}
^7K = {1 \over 2} 
\bigg( {\gamma_e \over ^7\gamma_n} \bigg) C
\sum_{i=1}^4   k({\bf r}_i)
\end{equation}
where 
\begin{equation}
k({\bf r}_i) = \sum_j \chi({\bf r}_i,{\bf r}_j)
\end{equation}
with $j$ summing over the whole lattice.
Note that for the pure system $k({\bf r}_i)$ reduces to 
$\chi_{pure}({\bf q}\rightarrow 0)$.

\vspace{0.4 in}

\begin{center}
{\bf {Appendix II: {\boldmath$V_{\rm eff}$} at lowest order in 
{\boldmath$U$}}}
\end{center}

It is possible that the effective impurity potential 
has an extended component because of the correlated nature 
of the host material \cite{Ziegler}.
At lowest order in the bare impurity potential $V_0$ 
and the Coluomb repulsion $U$, such a contribution 
originates from the scattering process shown in 
Fig.~15.
This process leads to a momentum-dependent effective interaction 
which is given by 
\begin{equation}
V^{(1)}_{\rm eff}({\bf q}) = - V_0 U 
\chi_0^L({\bf q}),
\end{equation}
where $\chi_0^L$ is the Lindhard susceptibility of the pure 
system, Eq.~(\ref{Lindhard}).
Since $\chi_0^L({\bf q})$ peaks at 
${\bf q}\sim (\pi,\pi)$, 
$V_{\rm eff}^{(1)}$ is attractive at sites neighboring the 
impurity when $V_0<0$.
For large $V_0$, it is necessary to replace the bare impurity 
potential in Fig.~15 by the impurity scattering $t$-matrix, 
in which case one obtains
\begin{eqnarray}
V^{(1)}_{\rm eff}({\bf q}) = U {T \over N}
\sum_{{\bf k},i\omega_{n'}} \,
G_0&&({\bf k},i\omega_{n'})  
G_0({\bf k}+{\bf q},i\omega_{n'}) \nonumber \\
&&
\times
{ V_0 \over 1 - V_0 F_0(i\omega_{n'}) },
\end{eqnarray}
where
\begin{equation}
F_0(i\omega_n) = {1 \over N} \sum_{\bf p}
G_0({\bf p},i\omega_n).
\end{equation}
The Fourier transform
\begin{equation}
V_{\rm eff}^{(1)}({\bf r}) = {1 \over N} \sum_{\bf q} 
e^{i {\bf q} \cdot {\bf r}} 
V_{\rm eff}^{(1)}({\bf q})
\end{equation}
gives the real-space structure of the effective interaction. 
This calculation of $V_{\rm eff}^{(1)}({\bf r})$ has been 
carried out for $V_0=-100t$, $T=100$~K and $W=3$~eV,
and it is found that $V_{\rm eff}^{(1)}({\bf r})$ 
at ${\bf r}=(1,0)$ is $-0.017t$.
Even though $V_{\rm eff}^{(1)}({\bf r}=(1,0))$
is found to be attractive in leading order in $U$, 
its magnitude is smaller by a factor of three from $V_1=-0.05t$
found by fitting the $^7K$ data on optimally doped 
YBa$_2$Cu$_3$O$_{6+x}$.

This is the behavior expected for $V_{\rm eff}$ at lowest 
order in $U$.
Clearly, an approximation which is first order in $U$ 
would be insufficient. 
The actual structure of $V_{\rm eff}$, which includes
scattering processes to all orders in $U$, depends on the 
reducible particle-particle vertex in the singlet and the 
triplet channels.
Hence, it would be useful to calculate 
$V_{\rm eff}$ using exact numerical techniques in the paramagnetic 
state of the Hubbard or the $t$-$J$ models.
However, it should be noted that the structure in 
the electron density $n({\bf r})$ does not necessarily 
reflect the structure in $V_{\rm eff}$, as it was seen in 
Fig.~11(c).
In Ref.~\cite{Ziegler}, $V_{\rm eff}$ was calculated 
for the $t$-$J$ model but at half-filling, where $V_{\rm eff}$ 
at the $(1,0)$ site was found to be repulsive.
However, the insulating state might be different than the 
paramagnetic state, which is considered here.
In Ref.~\cite{Bulut2000}, an effective impurity interaction 
which is repulsive at ${\bf r}=(1,0)$ was used 
for calculating the enhancement of the uniform susceptibility 
by dilute nonmagnetic impurities.
Since both positive and negative values of $V_1$ act to enhance 
the uniform susceptibility, 
this does not determine the sign of $V_1$.
On the other hand,
the Knight shift data studied here require that, 
at the level of RPA, 
$V_{\rm eff}$ at the nearest-neighbor sites of the impurity 
is weakly attractive.

\acknowledgments

The author gratefully acknowledges helpful discussions 
with H. Alloul, J. Bobroff, P. Bourges, C. Hammel, B. Keimer,
Y. Sidis, and C.P. Slichter. 
The author also thanks H. Alloul and J. Bobroff for helpful comments 
on the manuscript
and the Laboratoire de Physique des Solides at Orsay 
for its hospitality.
The numerical computations reported in this paper were performed 
at the Center for Information Technology at Ko\c{c} University.


\newpage

\begin{figure}
\begin{center}
\leavevmode
\epsfxsize=10cm 
\epsfysize=10cm 
\epsffile[100 180 550 630]{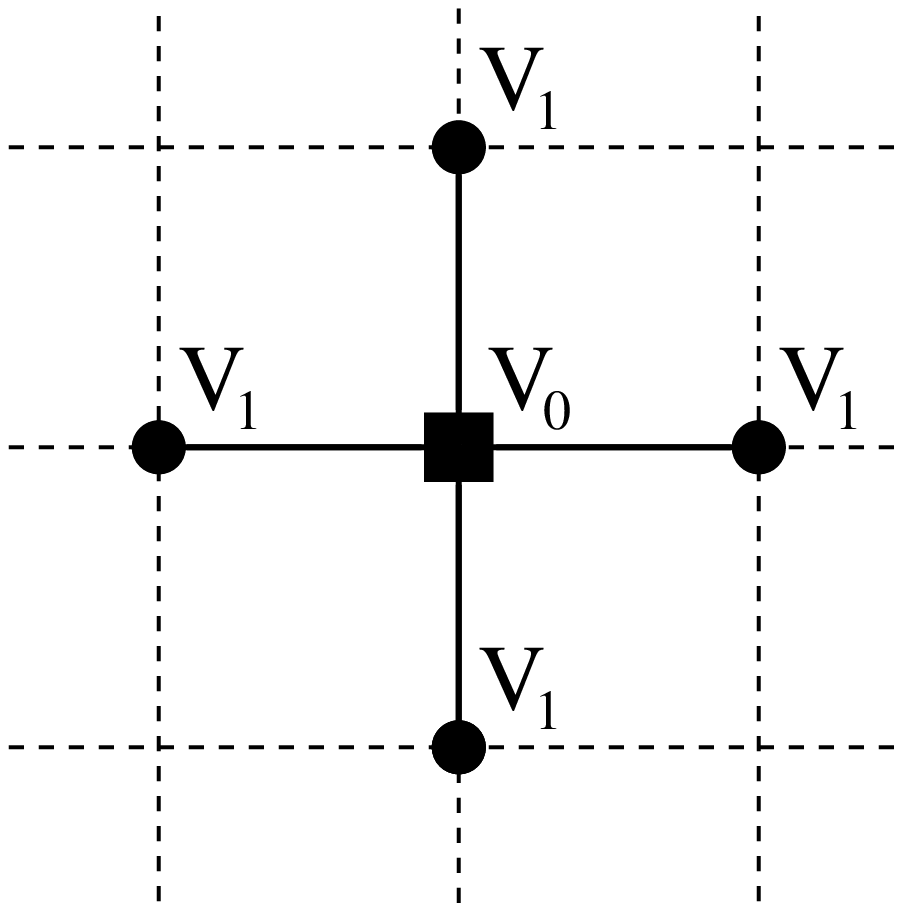}
\end{center}
\caption{
Illustration of the extended impurity potential due to 
a nonmagnetic impurity 
represented by the filled square at the center.
Here,
$V_0$ is the onsite component of the impurity potential and 
$V_1$ acts at the nearest-neighbour sites of the impurity.
}
\label{fig1}
\end{figure}

\newpage

\begin{figure} 
\begin{center}
\leavevmode
\epsfxsize=7.5cm 
\epsfysize=6.59cm 
\epsffile[100 170 550 580]{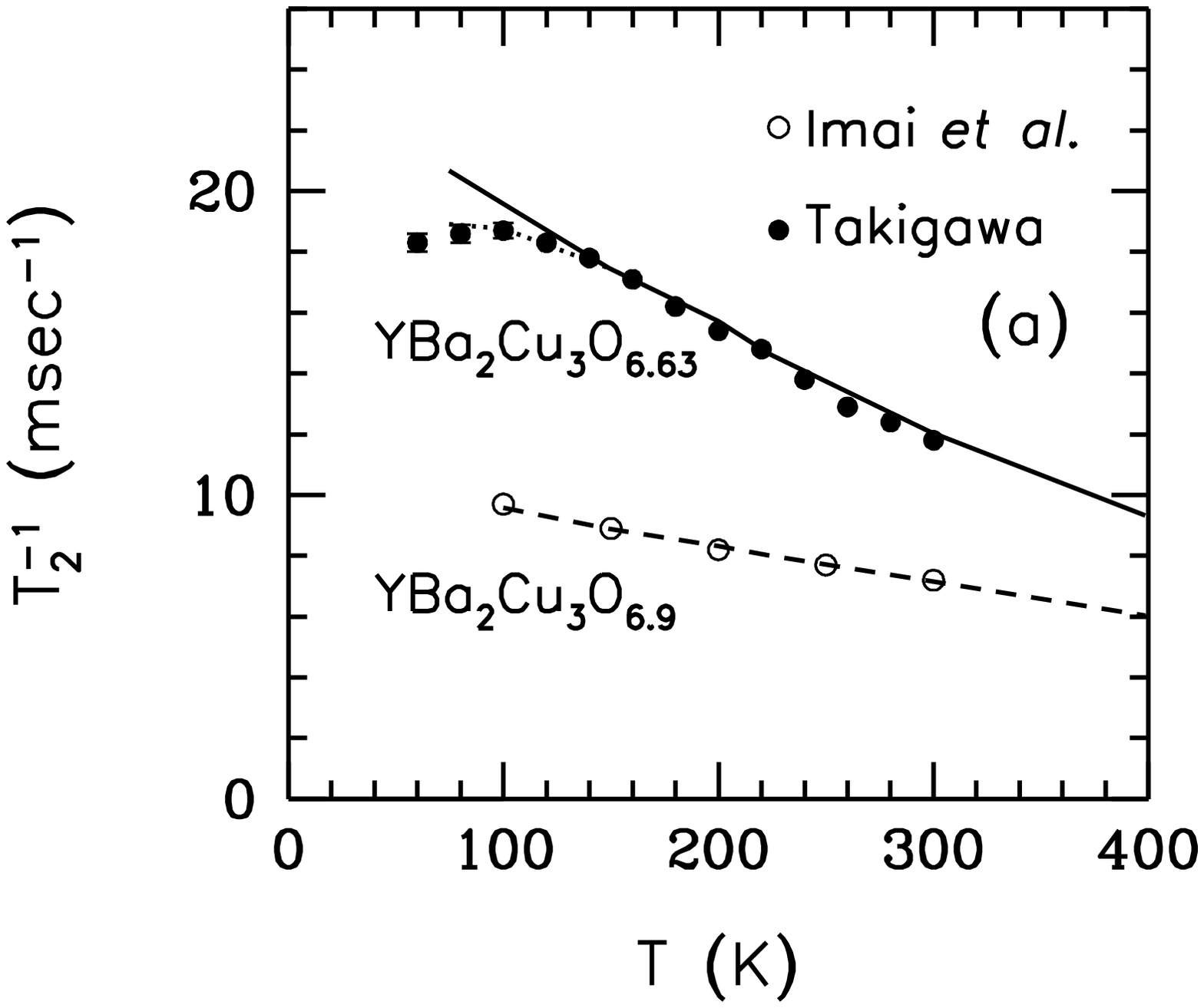}
\end{center}
\begin{center}
\leavevmode
\epsfxsize=7.5cm 
\epsfysize=6.59cm 
\epsffile[100 170 550 580]{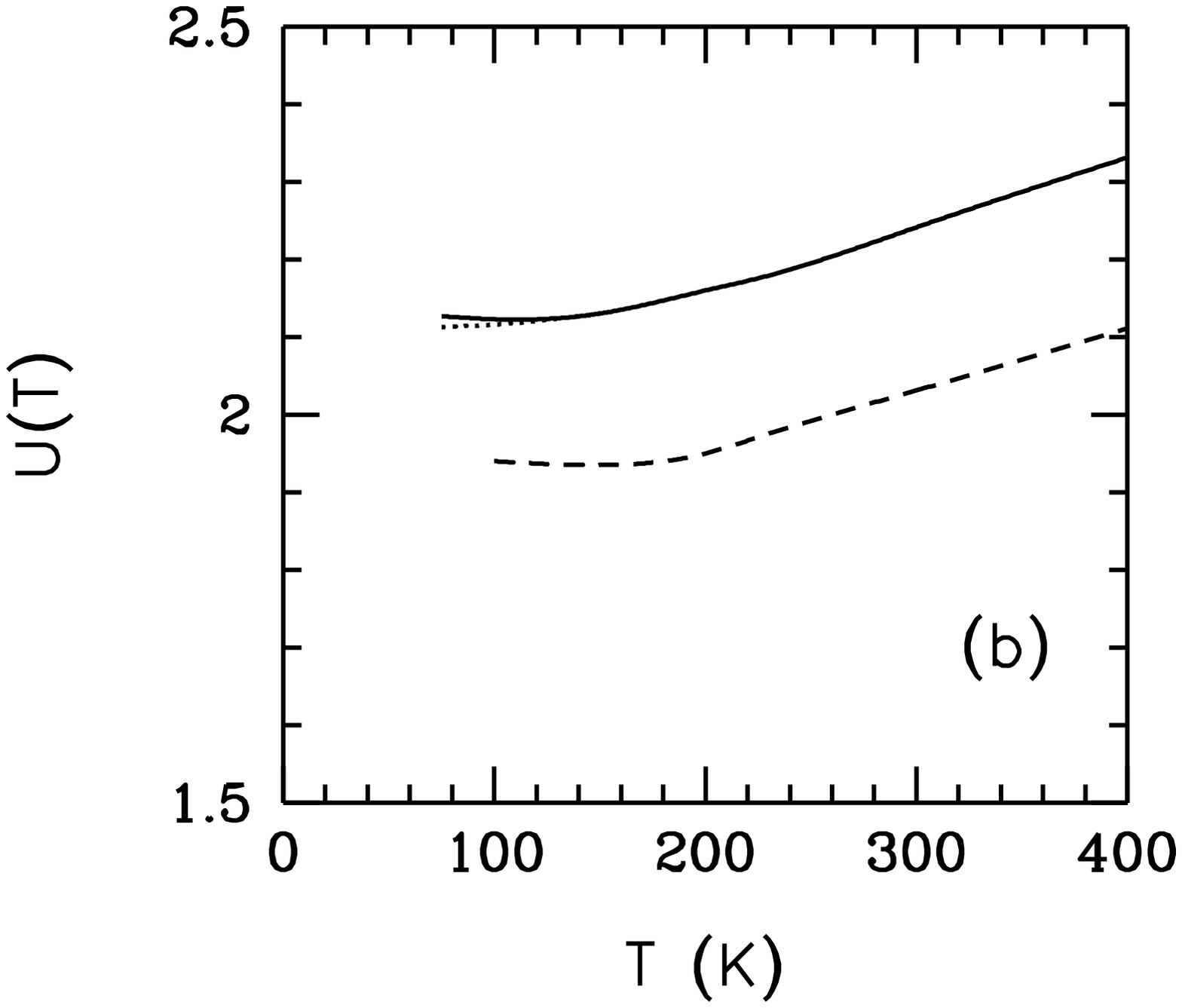}
\end{center}
\caption{
(a) Transverse nuclear relaxation rate $T_2^{-1}$ for 
$^{63}$Cu(2) versus $T$ in pure 
YBa$_2$Cu$_3$O$_{6+x}$. 
The open and the filled circles represent the experimental 
data by Imai {\it et al.} [24]
and Takigawa [25]
for 
YBa$_2$Cu$_3$O$_{6.9}$
and for 
YBa$_2$Cu$_3$O$_{6.63}$, 
respectively.
The dashed line is a fit of the $T_2^{-1}$ data 
by Imai {\it et al.} [24] on
YBa$_2$Cu$_3$O$_{6.9}$.
For 
YBa$_2$Cu$_3$O$_{6.63}$, 
the fitting of the data below 160~K
has been carried out in two ways. 
The solid line has been obtained by 
ignoring the 
saturation of $T_2^{-1}$ below $\sim 160$ K, 
and 
the dotted curve has been obtained by fitting the data for 
80~K$<T<160$~K.
(b) Temperature dependence of $U$.
The dashed curve has been obtained by fitting the $T_2^{-1}$ 
measurements on pure 
YBa$_2$Cu$_3$O$_{6.9}$.
The solid and the dotted curves have been obtained for pure
YBa$_2$Cu$_3$O$_{6.63}$ 
by fitting the solid and the dotted curves seen in (a).
}
\label{fig2}
\end{figure}

\newpage

\begin{figure} 
\begin{center}
\leavevmode
\epsfxsize=7.5cm 
\epsfysize=6.59cm 
\epsffile[100 170 550 580]{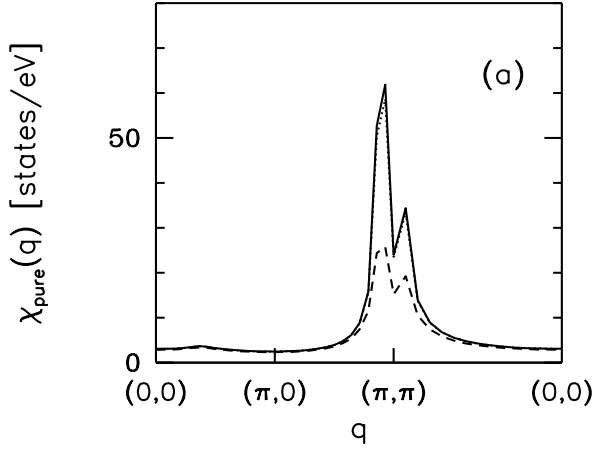}
\end{center}
\begin{center}
\leavevmode
\epsfxsize=7.5cm 
\epsfysize=6.59cm 
\epsffile[100 170 550 580]{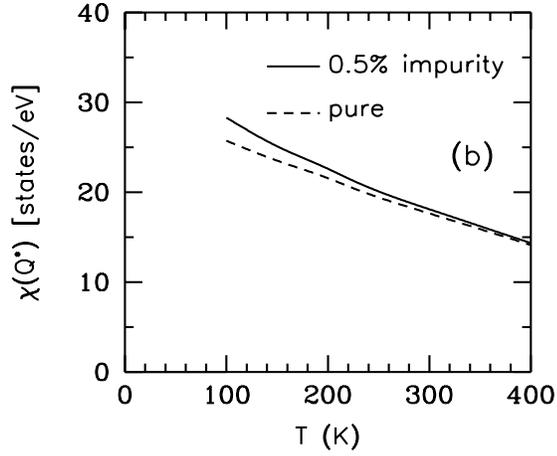}
\end{center}
\begin{center}
\leavevmode
\epsfxsize=7.5cm 
\epsfysize=6.59cm 
\epsffile[100 170 550 580]{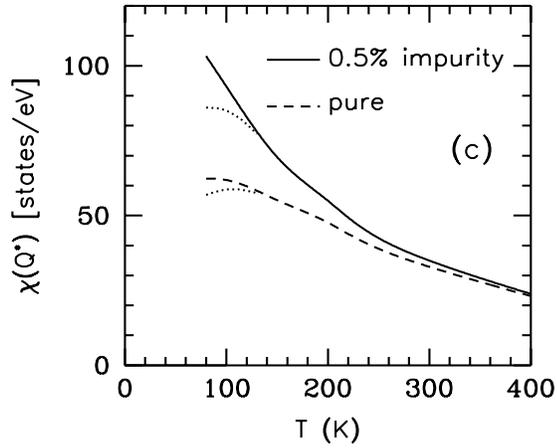}
\end{center}
\caption{
(a) $\chi_{pure}({\bf q})$ versus ${\bf q}$ at 100~K
for the pure system.
The dashed curve represents results for 
optimally doped
YBa$_2$Cu$_3$O$_{6+x}$.
The solid and the dotted curves have been obtained for 
underdoped
YBa$_2$Cu$_3$O$_{6+x}$.
(b) $\chi({\bf Q^*})$ versus $T$
for optimally doped 
YBa$_2$Cu$_3$O$_{6+x}$,
where ${\bf Q^*}$ is the wavevector at which $\chi({\bf q})$ peaks
at low temperatures.
Here, the dashed curve is for optimally doped pure
YBa$_2$Cu$_3$O$_{6+x}$, and 
the solid curve is for the 
case of 0.5\% dilute impurities.
(c) $\chi({\bf Q^*})$ versus $T$
for underdoped 
YBa$_2$Cu$_3$O$_{6+x}$.
The dashed curve is for the pure system 
and the solid curve is for 
the case with 0.5\% impurities.
The dotted curves were obtained by using the values of $U$
given by the dotted curves in Fig.~2(b).
The results for the impure systems 
shown in (b) and (c)
were calculated for $V_1=-0.15t$
as described in Section IV.A.
}
\label{fig3}
\end{figure}

\newpage

\begin{figure} 
\begin{center}
\leavevmode
\epsfxsize=10cm 
\epsfysize=10cm 
\epsffile[100 180 550 630]{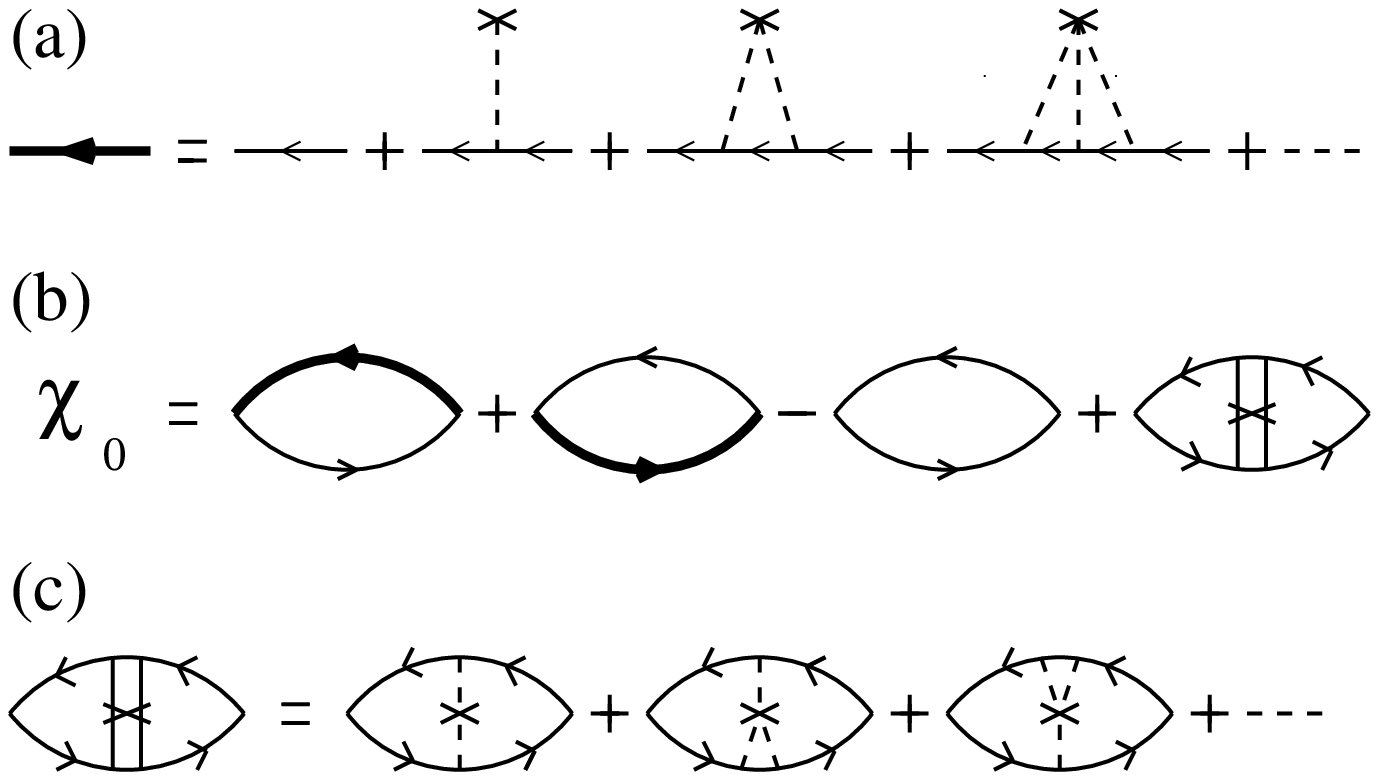}
\end{center}
\caption{
Feynman diagrams for 
(a) the dressed single-particle Green's function $G$ and 
(b)-(c) the irreducible susceptibility $\chi_0$
within the presence of one impurity.
}
\label{fig4}
\end{figure}

\newpage

\begin{figure} 
\begin{center}
\leavevmode
\epsfxsize=7.5cm 
\epsfysize=6.59cm 
\epsffile[100 200 550 610]{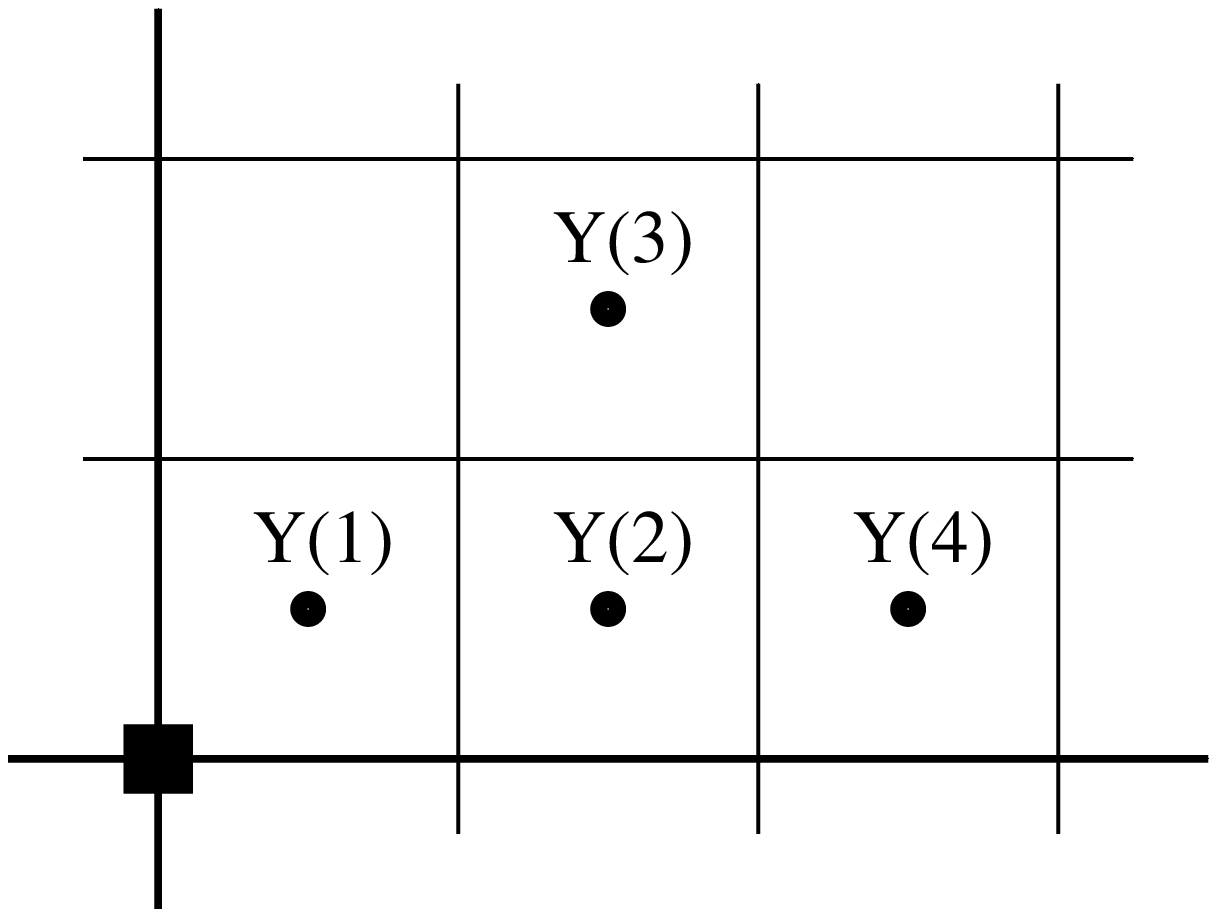}
\end{center}
\caption{
Sketch of the neighbouring Y($i$) sites of 
a nonmagnetic impurity located in the CuO$_2$ plane.
Here, the filled square indicates the impurity site.
}
\label{fig5}
\end{figure}

\newpage

\begin{figure} 
\begin{center}
\leavevmode
\epsfxsize=7.5cm 
\epsfysize=6.59cm 
\epsffile[100 170 550 580]{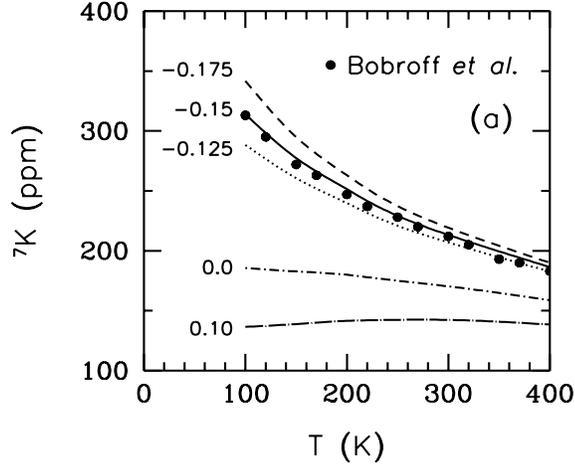}
\end{center}
\begin{center}
\leavevmode
\epsfxsize=7.5cm 
\epsfysize=6.59cm 
\epsffile[100 170 550 580]{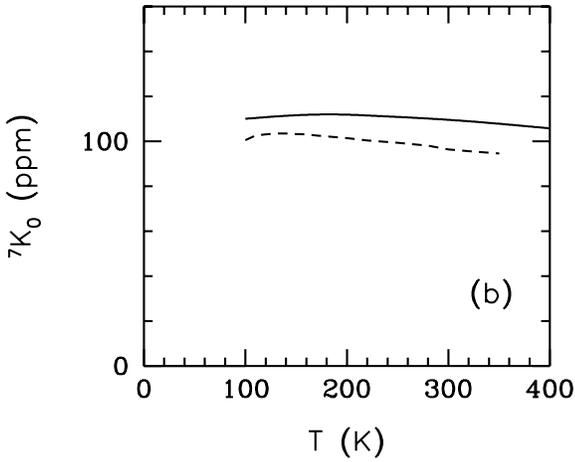}
\end{center}
\caption{
(a) $^7$Li Knight shift $^7K$ versus $T$ in
optimally doped 
YBa$_2$Cu$_3$O$_{6+x}$.
The filled circles represent the Knight shift data by Bobroff 
{\it et al.} [2].
The curves represent the results of the calculations 
for various values of $V_1/t$ 
which are indicated next to the curves.
(b) Estimated temperature dependence of 
$^7K_0$ for optimally doped 
YBa$_2$Cu$_3$O$_{6+x}$. 
Here, $^7K_0$ represents the value of the 
$^7$Li Knight shift, if the substitution of the Li impurity 
had not induced any changes in the magnetic correlations 
around it.
The dashed curve is the experimental estimate of 
$^7K_0$ obtained by using the $^{89}K$ data [38] on pure
optimally doped 
YBa$_2$Cu$_3$O$_{6+x}$. 
The solid line is the result obtained by setting 
$V_0=V_1=0$ in this model.
}
\label{fig6}
\end{figure}

\newpage

\begin{figure} 
\begin{center}
\leavevmode
\epsfxsize=7.5cm 
\epsfysize=6.59cm 
\epsffile[100 170 550 580]{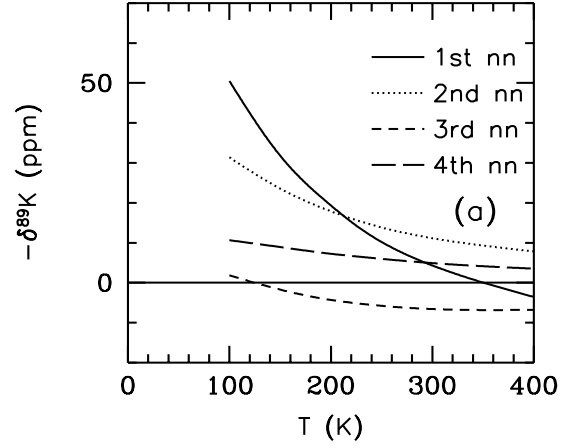}
\end{center}
\begin{center}
\leavevmode
\epsfxsize=7.5cm 
\epsfysize=6.59cm 
\epsffile[100 170 550 580]{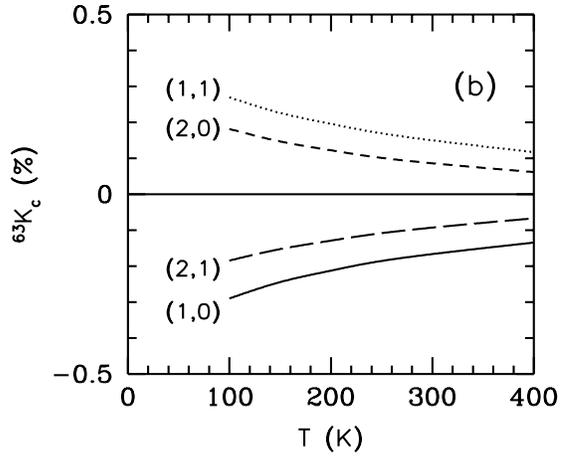}
\end{center}
\caption{
(a) Temperature dependence of $-\delta^{89}K$ 
at the sites which are the 1st, 2nd, 3rd 
and the 4th nearest
$^{89}$Y neighbors of the impurity in 
optimally doped
YBa$_2$Cu$_3$O$_{6+x}$.
Here,
$\delta^{89}K$ is the induced change in the 
$^{89}$Y Knight shift up on the substitution of the 
impurity.
(b) Temperature dependence of the $^{63}$Cu(2) 
Knight shift for ${\bf H}||{\bf c}$,
$^{63}K_c({\bf r}_i)$, 
at various sites ${\bf r}_i$,
which are indicated next to the curves,
for the optimally doped 
YBa$_2$Cu$_3$O$_{6+x}$.
The results shown in (a) and (b) 
were obtained by using 
$V_1= -0.15t$.
}
\label{fig7}
\end{figure}

\newpage

\begin{figure} 
\begin{center}
\leavevmode
\epsfxsize=7.5cm 
\epsfysize=6.59cm 
\epsffile[100 170 550 580]{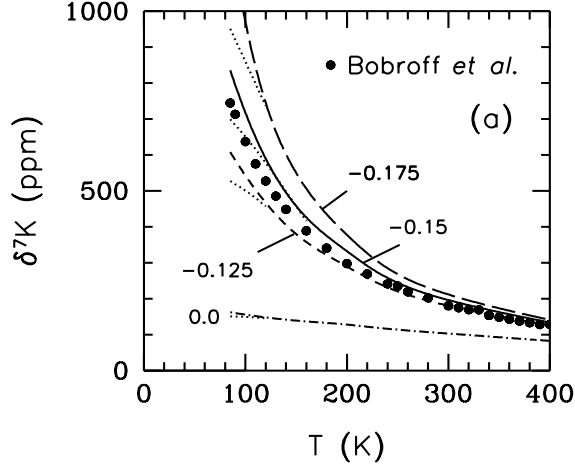}
\end{center}
\begin{center}
\leavevmode
\epsfxsize=7.5cm 
\epsfysize=6.59cm 
\epsffile[100 170 550 580]{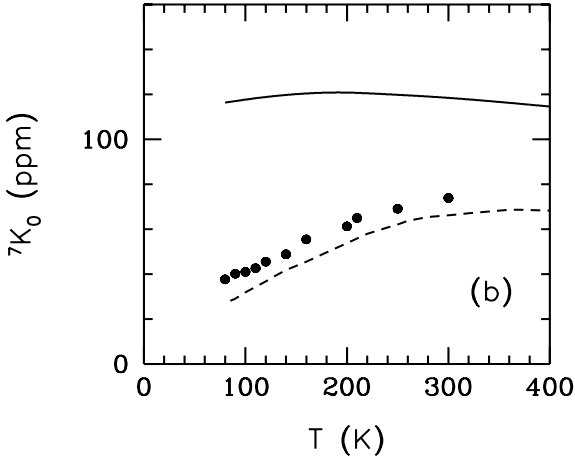}
\end{center}
\caption{
(a) Temperature dependence of 
$\delta^7K$
for underdoped 
YBa$_2$Cu$_3$O$_{6+x}$.
Here, $\delta^7K$ is defined as $^7K-^7K_0$, 
where $^7K$ is the $^7$Li Knight shift and 
$^7K_0$ represents the value of the $^7$Li Knight shift, 
if the substitution of the Li impurity had not changed 
the magnetic correlations around it. 
The filled circles represent the data 
by Bobroff {\it et al.} [2] on 
YBa$_2$Cu$_3$O$_{6.6}$.
The curves represent the results of the calculations 
for various values of $V_1/t$ which are
indicated next to the curves.
(b) Temperature dependence of 
$^7K_0$ for underdoped
YBa$_2$Cu$_3$O$_{6+x}$.
The solid points have been obtained by using the 
$^{89}K({\rm main})$ data on underdoped 
YBa$_2$Cu$_3$O$_{6+x}$ 
with Zn impurities [1,4],
while the dashed curve has been obtained by using the
$^{89}K$ data on pure underdoped 
YBa$_2$Cu$_3$O$_{6+x}$ [38].
The solid curve shows the results of the calculations for 
$V_0=V_1=0$.
}
\label{fig8}
\end{figure}

\newpage

\begin{figure} 
\begin{center}
\leavevmode
\epsfxsize=7.5cm 
\epsfysize=6.59cm 
\epsffile[100 170 550 580]{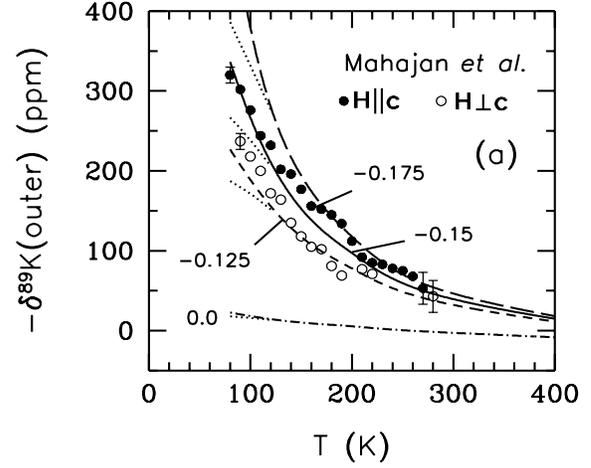}
\end{center}
\begin{center}
\leavevmode
\epsfxsize=7.5cm 
\epsfysize=6.59cm 
\epsffile[100 170 550 580]{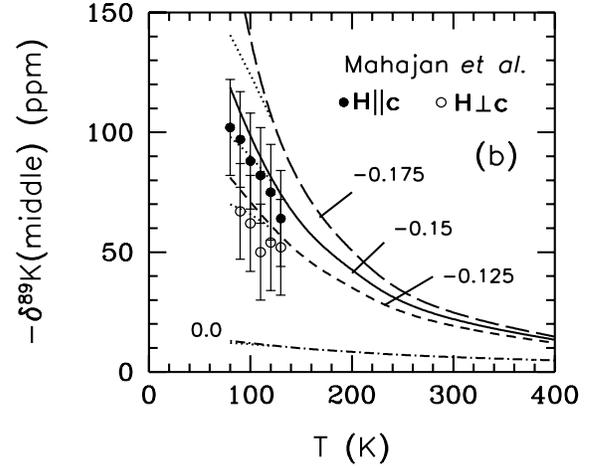}
\end{center}
\caption{
(a) Temperature dependence of $-\delta^{89}K(\rm outer)$
in underdoped 
YBa$_2$Cu$_3$O$_{6+x}$
with Zn impurities.
Here, $\delta^{89}K(\rm outer)$ is defined as the shift of the outer 
$^{89}$Y satellite with respect to the $^{89}$Y main resonance line.
The open and the filled circles represent the experimental data on 
$-\delta^{89}K({\rm outer})$ in underdoped
YBa$_2$Cu$_3$O$_{6+x}$
with Zn impurities
for the magnetic field ${\bf H}|| {\bf c}$ 
and ${\bf H}\perp {\bf c}$, 
respectively [1,4].
The curves represent the results obtained 
by using various values of $V_1/t$
which are indicated next to the curves 
for the Y(1) site, which is the nearest $^{89}$Y neighbour 
of the nonmagnetic impurity.
(b) Results similar to those in (a) but for $-\delta^{89}K(\rm middle)$.
Here, $\delta^{89}K(\rm middle)$ is defined as the shift of the middle 
$^{89}$Y resonance line with respect to the main line.
The curves represent results 
for various values of $V_1/t$ 
for the Y(2) site, which is the second-nearest
$^{89}$Y neighbour of the nonmagnetic impurity.
}
\label{fig9}
\end{figure}

\newpage

\begin{figure}
\begin{center}
\leavevmode
\epsfxsize=7.5cm 
\epsfysize=6.59cm 
\epsffile[100 170 550 580]{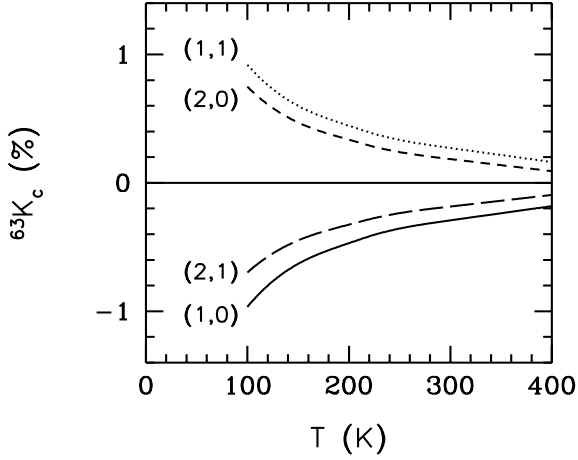}
\end{center}
\caption{
Temperature dependence of the $^{63}$Cu(2) Knight shift 
for ${\bf H}||{\bf c}$, $^{63}K_c({\bf r}_i)$, 
in underdoped 
YBa$_2$Cu$_3$O$_{6+x}$
at sites ${\bf r}_i$
which are indicated next to the curves.
These results were obtained for 
$V_1=-0.15t$ using the values of $U(T)$ given by the 
solid curve in Fig.~2(b).
}
\label{fig10}
\end{figure}


\begin{figure} 
\begin{center}
\leavevmode
\epsfxsize=7.5cm 
\epsfysize=6.59cm 
\epsffile[100 170 550 580]{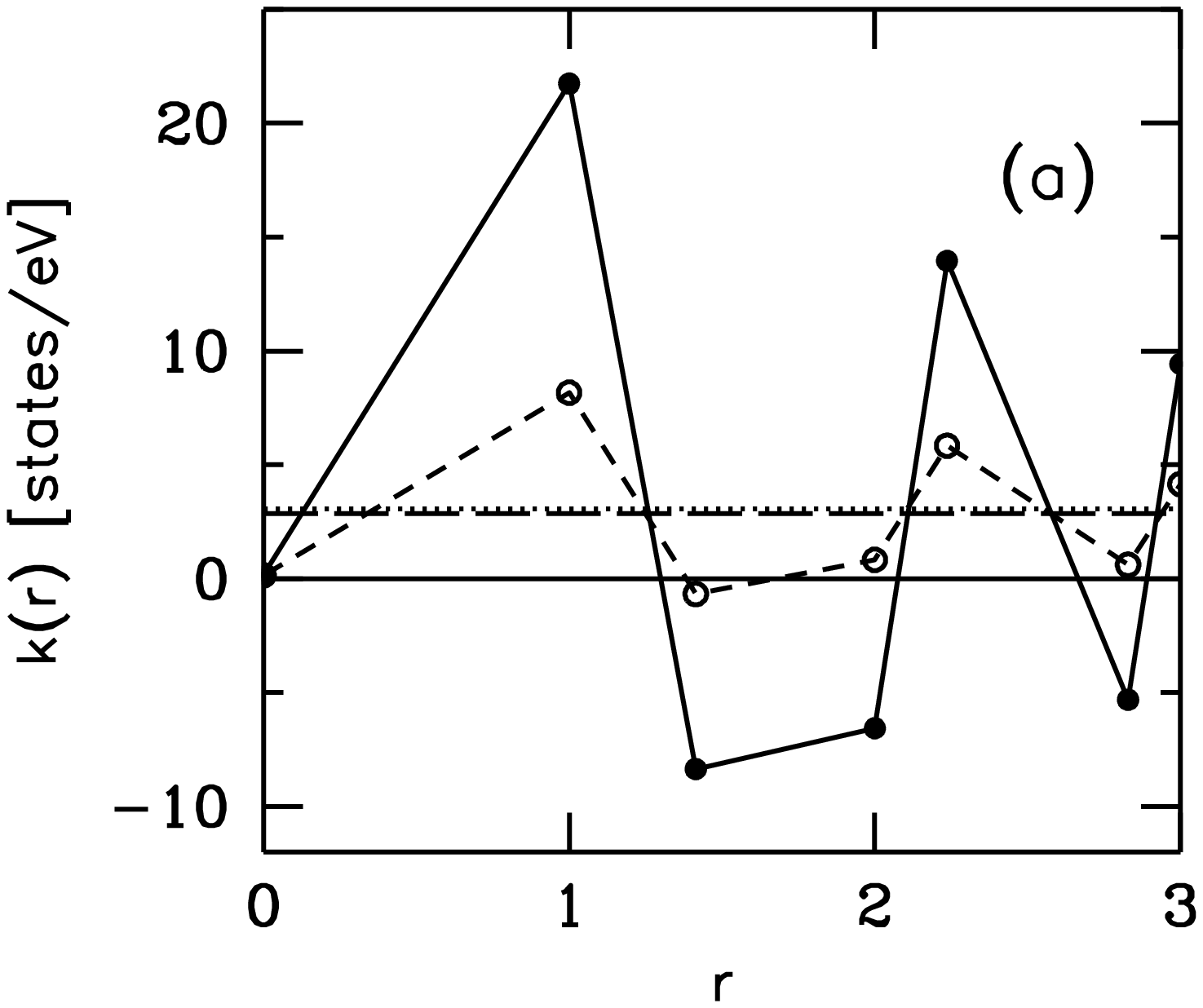}
\end{center}
\begin{center}
\leavevmode
\epsfxsize=7.5cm 
\epsfysize=6.59cm 
\epsffile[100 170 550 580]{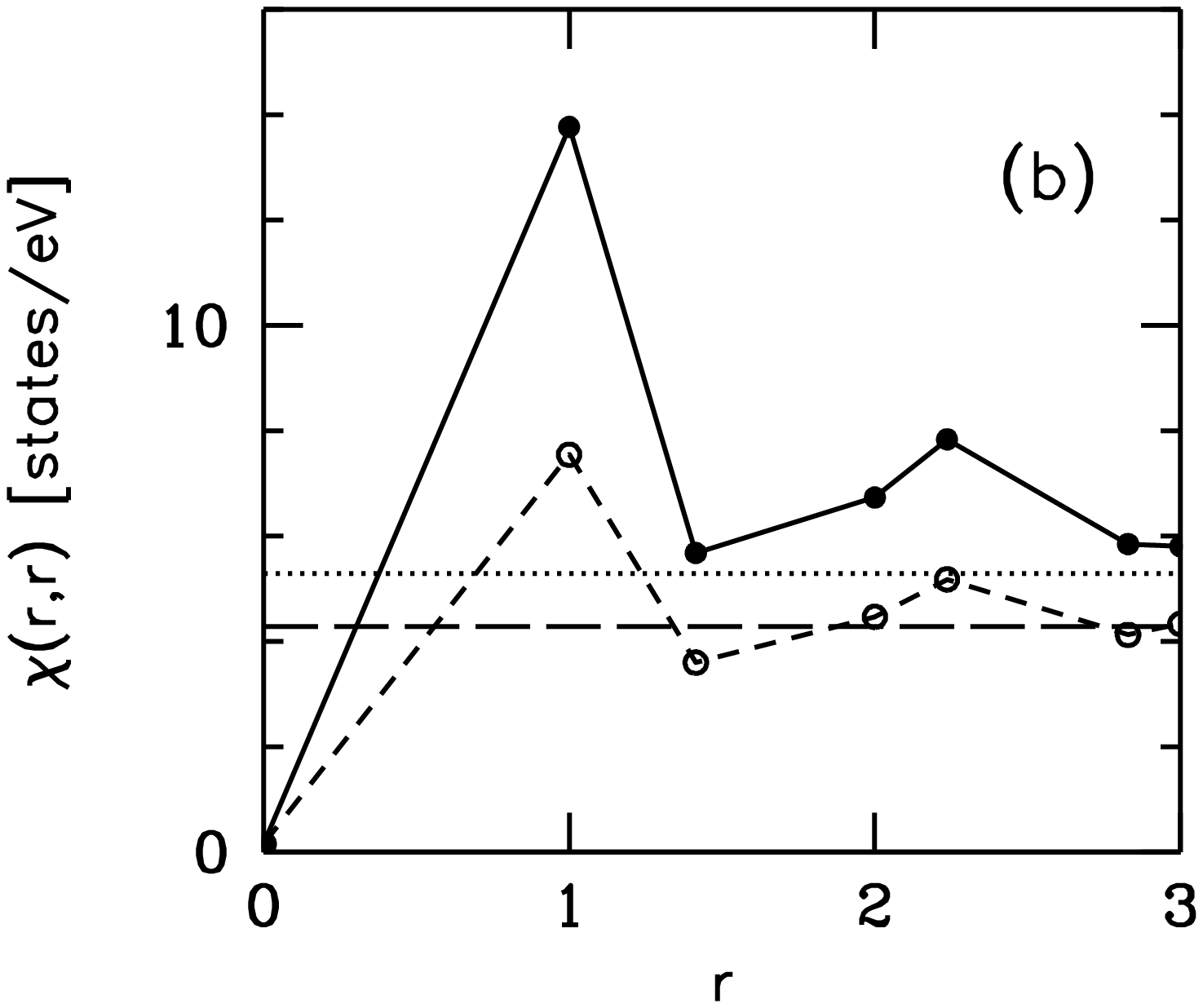}
\end{center}
\begin{center}
\leavevmode
\epsfxsize=7.5cm 
\epsfysize=6.59cm 
\epsffile[100 170 550 580]{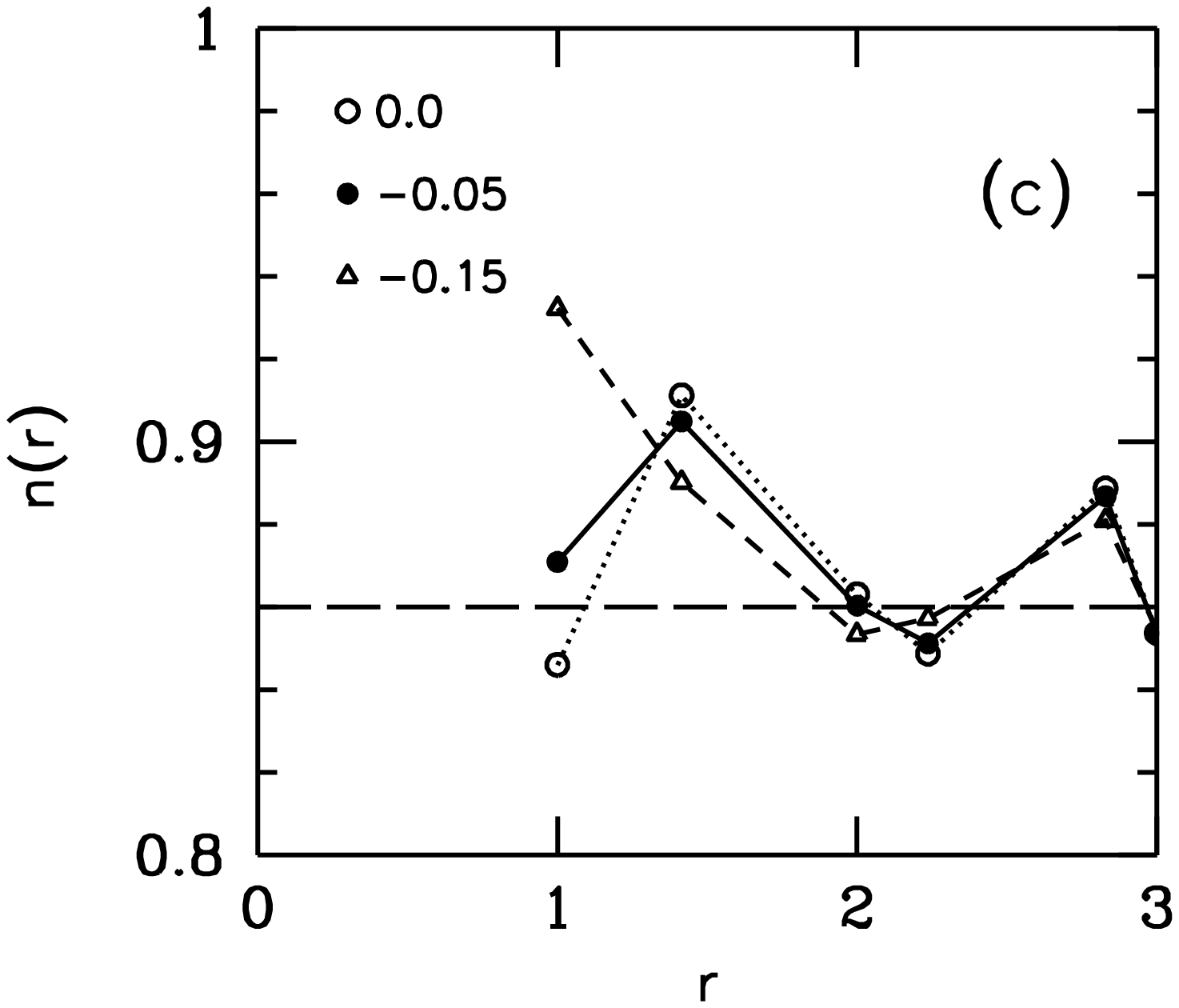}
\end{center}
\caption{
(a) $k({\bf r})$ versus 
the distance $r$ away from the impurity 
in units of the lattice spacing $a$ for $T=100$~K
and $V_1=-0.15t$.
Here, the open and the filled circles represent 
the results for the optimally doped and the underdoped
YBa$_2$Cu$_3$O$_{6+x}$,
respectively.
Also,
the horizontal long-dashed and dotted lines represent 
$k({\bf r})$ for the 
optimally doped and underdoped pure systems, respectively.
(b) $\chi({\bf r},{\bf r})$ versus the distance $r$
away from the impurity presented in the same way as in 
(a) for $k({\bf r})$.
(c) Electron occupation $n({\bf r}_i)$ at sites ${\bf r}_i$ near 
the impurity plotted as a function 
$r=|{\bf r}_i|$ at 100 K.
These results were obtained for $V_1/t=-0.15$ with 
$W=1$ eV (solid) and for $V_1/t=-0.05$ with $W=3$ eV (dashed).
Also shown are results for an onsite impurity potential ($V_1=0$) 
and $W=1$ eV (dotted).
In all of these cases, $V_0=-100t$ was used.
}
\label{fig11}
\end{figure}

\newpage

\begin{figure} 
\begin{center}
\leavevmode
\epsfxsize=7.5cm 
\epsfysize=6.59cm 
\epsffile[100 170 550 580]{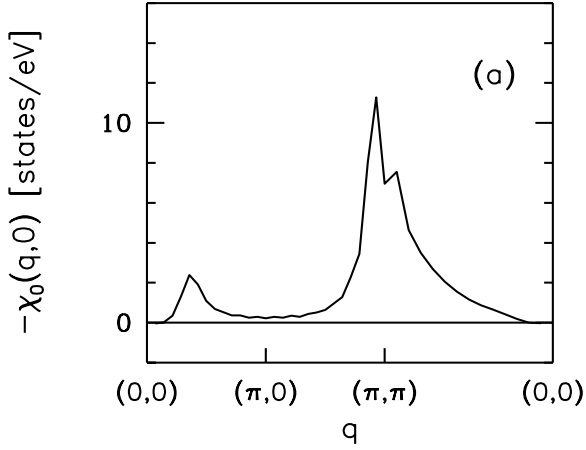}
\end{center}
\begin{center}
\leavevmode
\epsfxsize=7.5cm 
\epsfysize=6.59cm 
\epsffile[100 170 550 580]{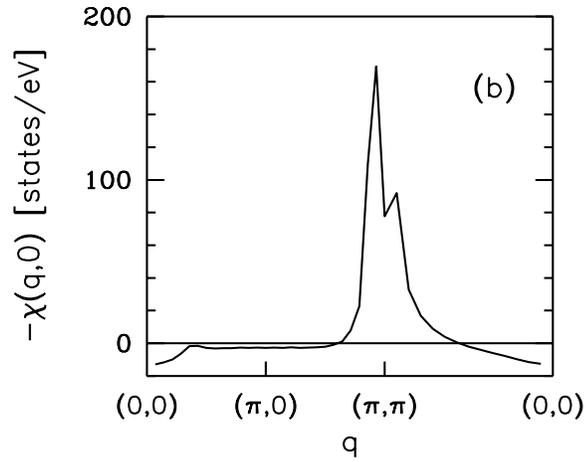}
\end{center}
\caption{
Off-diagonal susceptibilities 
(a) $-\chi_0({\bf q},{\bf q'}=0)$ and 
(b) $-\chi({\bf q},{\bf q'}=0)$ versus ${\bf q}$
obtained by using $V_1=-0.15t$ for 
optimally doped 
YBa$_2$Cu$_3$O$_{6+x}$ 
at $T=100$ K.
Here, the $\delta$-function components at 
${\bf q}={\bf q}'=0$ have been omitted.
}
\label{fig12}
\end{figure}

\newpage

\begin{figure} 
\begin{center}
\leavevmode
\epsfxsize=7.5cm 
\epsfysize=6.59cm 
\epsffile[100 170 550 580]{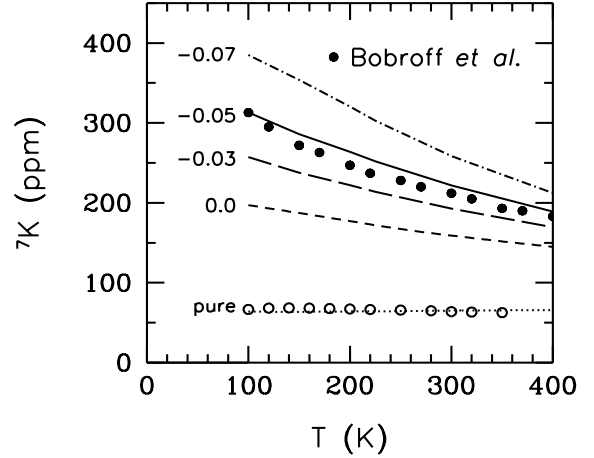}
\end{center}
\caption{
Temperature dependence of $^7K$ for optimally doped 
YBa$_2$Cu$_3$O$_{6+x}$, 
obtained for various values of $V_1/t$.
These results are for an effective bandwidth of 3 eV,
and here the $^7$Li hyperfine coupling $C$
was taken to be 1.4 kOe/$\mu_B$.
The filled circles represent the $^7K$ data on optimally doped
YBa$_2$Cu$_3$O$_{6+x}$
by Bobroff {\it et al.} [2].
The open circles represent the estimate of 
$^7K_0$ obtained by using the $^{89}K$ data [38] on pure 
optimally doped 
YBa$_2$Cu$_3$O$_{6+x}$.
Here, the $^{89}$Y hyperfine coupling was taken to be 
-2~kOe/$\mu_B$.
The dotted curve has been obtained by using $V_0=V_1=0$ and the
remaining curves have been obtained by using $V_0=-100t$ 
and the values of $V_1/t$ 
which are indicated next to the curves.
}
\label{fig13}
\end{figure}

\newpage

\begin{figure} 
\begin{center}
\leavevmode
\epsfxsize=7.5cm 
\epsfysize=6.59cm 
\epsffile[100 170 550 580]{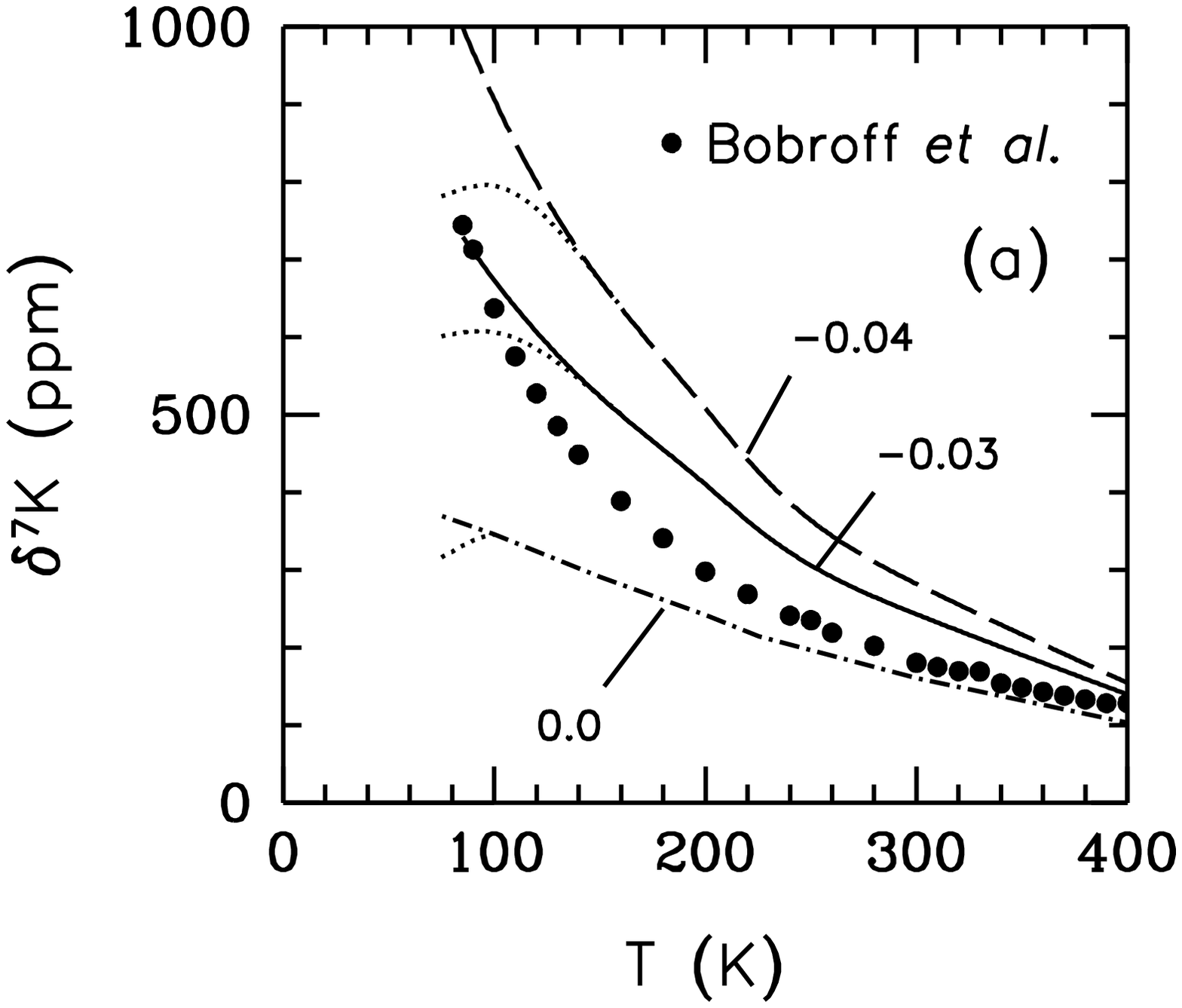}
\end{center}
\begin{center}
\leavevmode
\epsfxsize=7.5cm 
\epsfysize=6.59cm 
\epsffile[100 170 550 580]{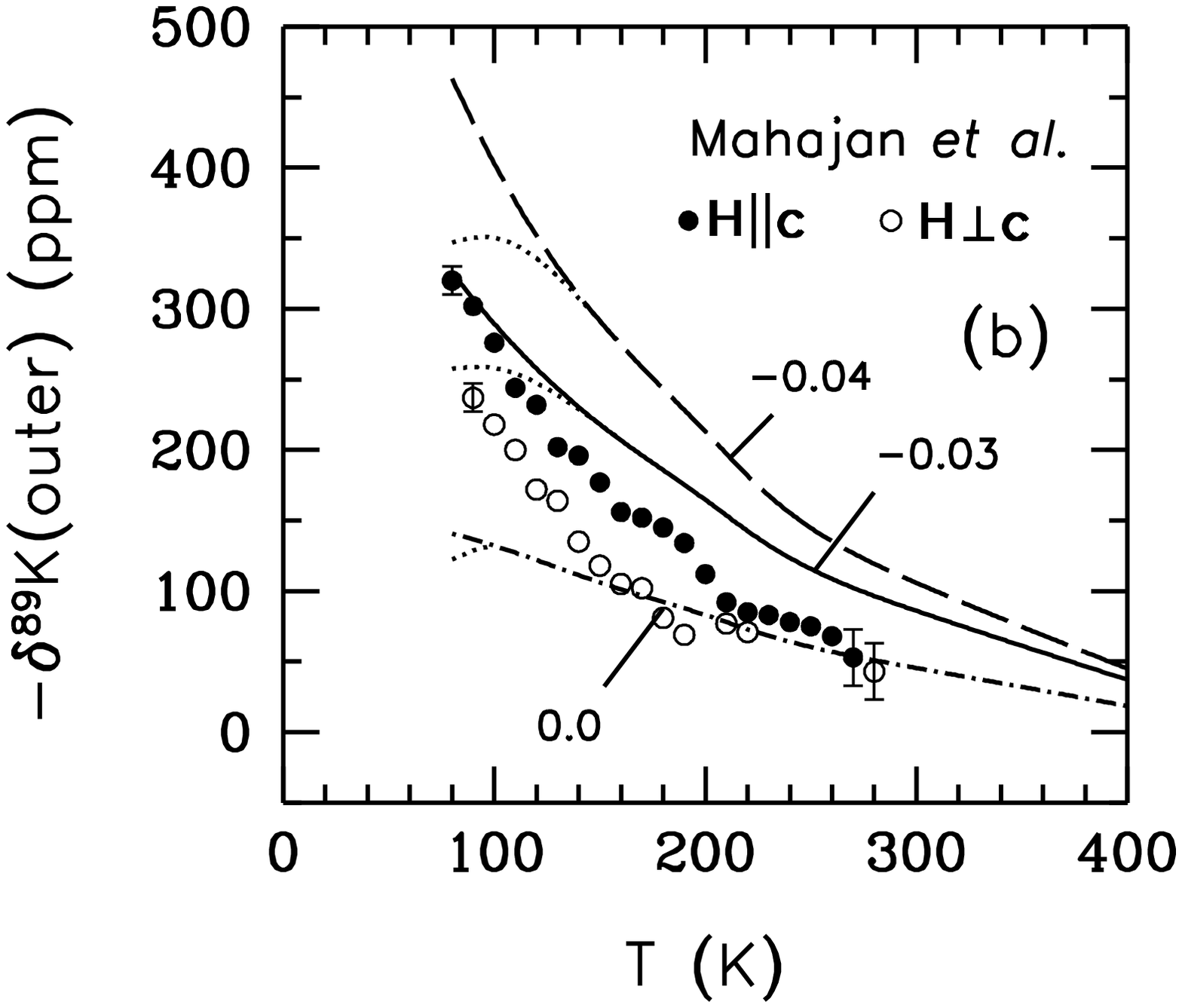}
\end{center}
\begin{center}
\leavevmode
\epsfxsize=7.5cm 
\epsfysize=6.59cm 
\epsffile[100 170 550 580]{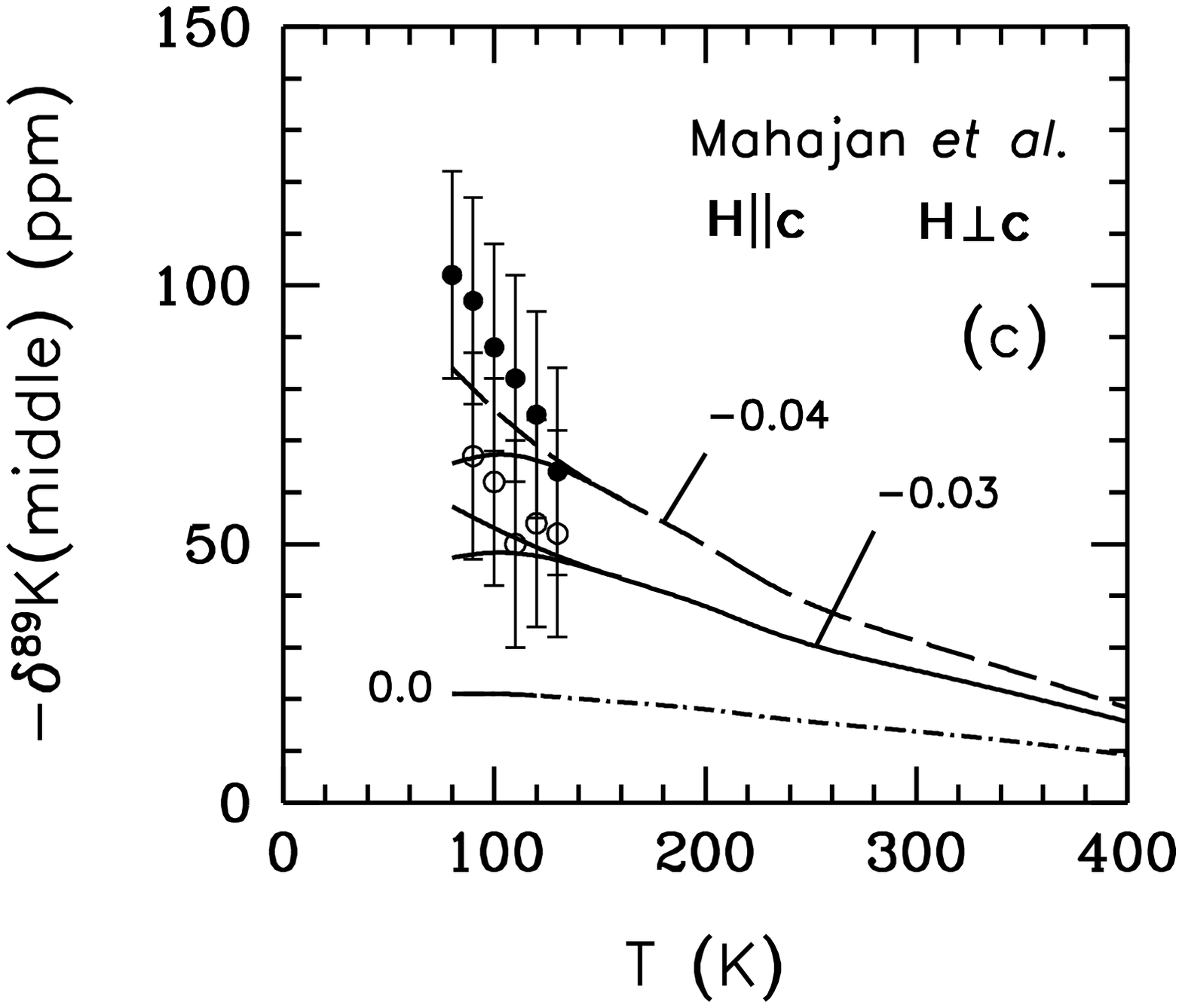}
\end{center}
\caption{
Temperature dependence of 
(a) $\delta^7K$, (b) $-\delta^{89}K$(outer) and 
(c) $-\delta^{89}K$(middle) for underdoped 
YBa$_2$Cu$_3$O$_{6+x}$
obtained by using various values of $V_1/t$
which are indicated next to the curves. 
These results have been calculated 
by using an effective bandwidth $W$ of 3~eV,
as compared to $W=1$~eV used in Figs.~8 and 9.
In addition, here the $^7$Li and the $^{89}$Y hyperfine couplings 
were taken to be 1.4~kOe/$\mu_B$ and 
-2~kOe/$\mu_B$, respectively. 
The experimental data in (a) are from Ref.~[2], and the data 
in (b) and (c) are from Refs.~[1,4].
}
\label{fig14}
\end{figure}

\newpage

\begin{figure} 
\begin{center}
\leavevmode
\epsfxsize=7.5cm 
\epsfysize=6.59cm 
\epsffile[100 200 550 610]{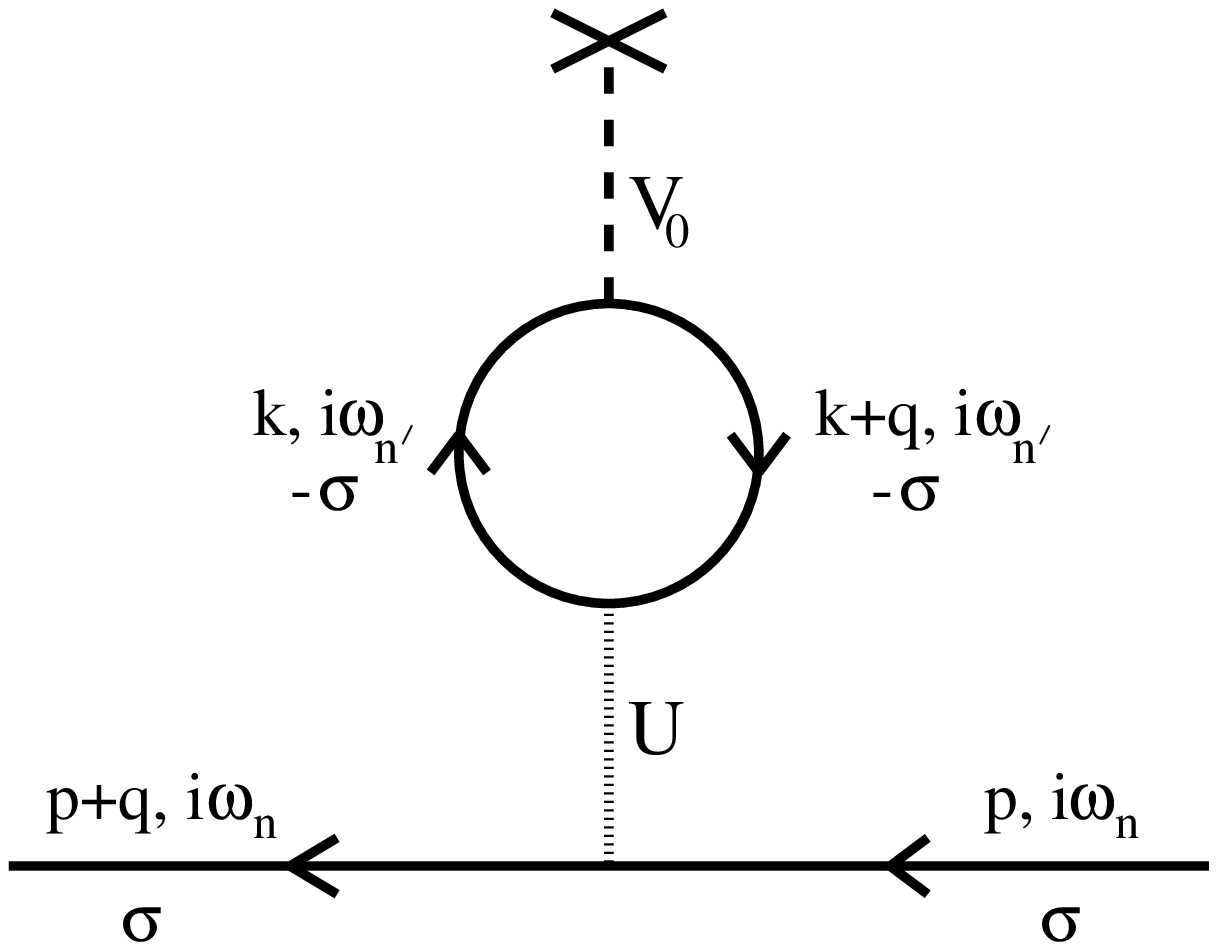}
\end{center}
\caption{
Feynman diagram for a scattering process which causes the effective 
impurity potential to be extended in real space.
This diagram is lowest order in $U$ and 
the bare impurity potential $V_0$.
Here, an electron with momentum ${\bf p}$, Matsubara frequency 
$\omega_n$ and spin $\sigma$ scatters to a state with 
momentum ${\bf p+q}$, Matsubara frequency $\omega_n$ and 
spin $\sigma$.
}
\label{fig15}
\end{figure}


\begin{thebibliography}{999}
\bibitem{89Y} When the magnetic resonance spectrum of 
$^{89}$Y in underdoped 
YBa$_2$Cu$_3$O$_{6+x}$
with Zn impurities is compared with the 
$^{89}$Y spectrum within the presence of Li
impurities, it is found that
$^{89}K({\rm outer})$ and 
$^{89}K({\rm middle})$ are shifted by about 20\% more to higher values 
for Li compared to Zn \cite{Bobroff}.
This could be because the stoichiometry 
of the Zn substituted and the Li substituted samples 
might be different \cite{Bobroff}.
For this reason, it would have been probably better to divide 
the calculated values of $\delta^{89}K$'s by a factor of 1.2
before comparing them with the 
$\delta ^{89}K$ experimental data.
However, this will not qualitatively affect the fits 
seen in Figs.~9(a) and (b).
\bibitem{Li7} Comparisons with the $^7$Li $T_1^{-1}$ data 
will also be useful in estimating the magnitude of the 
$^7$Li hyperfine coupling.
\end{thebibliography}
\end{document}